\documentclass[fleqn,usenatbib]{mnras}

\usepackage{newtxtext,newtxmath}

\usepackage[T1]{fontenc}

\DeclareRobustCommand{\VAN}[3]{#2}
\DeclareUnicodeCharacter{2212}{-}
\let\VANthebibliography\thebibliography
\def\thebibliography{\DeclareRobustCommand{\VAN}[3]{##3}\VANthebibliography}

\usepackage{graphicx}	
\usepackage{amsmath}	
\usepackage{ulem}
\usepackage{rotating}
\usepackage{xcolor}

\newcommand{\Msun}{\mbox{M$_{\odot}$}}

\newcommand{\Teff}{\mbox{$T_{\mathrm{eff}}$}}

\newcommand{\gppr}{\stackrel{>}{\scriptstyle \sim}}
\newcommand{\gappr}{\raisebox{-0.4ex}{$\gppr$}}

\title[Spin periods of magnetic white dwarfs]{Rotation plays a role in the generation of magnetic fields in single white dwarfs}

\author[Hernandez et al.]{
Mercedes S. Hernandez$^{1,2}$\thanks{E-mail: mercedes.hernandez@usm.cl},
Matthias R. Schreiber$^{1,2}$\thanks{E-mail: matthias.schreiber@usm.cl},
John D. Landstreet,$^{3,4}$ 
Stefano Bagnulo$^{3}$, 
\newauthor
Steven G. Parsons$^{5}$,
Martin Chavarria$^{1,2}$,
Odette Toloza$^{1,2}$,
Keaton J. Bell$^{6}$.
\\
\\
$^{1}$Departamento de F{\'i}sica, Universidad T\'ecnica Federico Santa Mar\'ia, Av. España 1680, Valpara{\'i}so, Chile.\\
$^{2}$Millennium Nucleus for Planet Formation, NPF, Valpara{\'i}so, 2340000, Chile.\\
$^{3}$Armagh Observatory and Planetarium, College Hill, Armagh BT61 9DG, Northern Ireland, UK.\\
$^{4}$University of Western Ontario, Department of Physics $\&$ Astronomy, London, ON N6A 3K7, Canada.\\
$^{5}$Department of Physics and Astronomy, University of Sheffield, Sheffield S3 7RH, UK.\\
$^{6}$Department of Physics, Queens College, City University of New York, Flushing, NY-11367, USA.
}

\date{Accepted 2024 January 25. Received 2024 January 23; in original form 2023 May 24}

\pubyear{2022}

\begin{document}
\label{firstpage}
\pagerange{\pageref{firstpage}--\pageref{lastpage}}
\maketitle

\begin{abstract}
 
%
%
Recent surveys of close white dwarf binaries as well as single white dwarfs have provided evidence for the late appearance of magnetic fields in white dwarfs, and a possible generation mechanism a crystallization and rotation-driven dynamo has been suggested. 
A key prediction of this dynamo is that magnetic white dwarfs rotate, at least on average, faster than their non-magnetic counterparts and/or that the magnetic field strength increases with rotation. Here we present rotation periods of ten white dwarfs  within 40\,pc measured using photometric variations. Eight of the light curves come from {\it TESS} observations and are thus not biased towards short periods, in contrast to most period estimates that have been reported previously in the literature. 
These {\it TESS} spin periods are indeed systematically shorter than those of non-magnetic white dwarfs.
This means that the crystallization and rotation-driven dynamo could be responsible for a fraction of the magnetic fields in white dwarfs. However, the full sample of magnetic white dwarfs also contains slowly rotating strongly magnetic white dwarfs which indicates that another mechanism that leads to the late appearance of magnetic white dwarfs might be at work, either in addition to or instead of the dynamo. The fast-spinning and massive magnetic white dwarfs 
that appear in the literature form a small fraction of magnetic white dwarfs, and probably result from a channel related to  white dwarf mergers. 
\end{abstract}




\section{Introduction}

White dwarfs have been speculated to have strong ($>1$\,MG) magnetic fields many 
decades ago when \citet{blackett47-1} postulated that the presumed fast rotation of white dwarfs can drive a dynamo, but the first detection of a magnetic white dwarf was obtained more than twenty years later \citep{kempetal70-1}. 
Ever since this groundbreaking discovery, the question of why some white dwarfs become strongly magnetic, while others do not, represents an unsolved questions of stellar evolution.

Today we know large numbers of magnetic white dwarfs. One of the most puzzling facts is the different fractions of strongly magnetic white dwarfs among single stars and in different binary star settings. 
The volume-limited sample of magnetic single white dwarfs established by \citet{bagnulo+landstreet21-1} shows that the fraction of magnetic white dwarfs increases with age and is about 20 per cent. 
Studies of the (incomplete) local 40\,pc sample of white dwarfs shows that massive white dwarfs (which are absent in the 20\,pc sample) often exhibit strong magnetic fields during the initial stages of the cooling phase
\citep{bagnulo+landstreet22-1}.

A high incidence of magnetic white dwarfs, i.e. around 36 per cent, is seen among cataclysmic variables (CVs), semi-detached close binary stars in which a white dwarf accretes from a Roche-lobe filling
main sequence star \citep{palaetal20-1}. 

Intriguingly, among the progenitors of CVs, detached white dwarf plus main-sequence star binaries, the fraction of systems with strongly magnetic white dwarfs is negligible 
\citep{liebertetal05-1,liebertetal15-1}. 
Of the more than one thousand known systems \citep{schreiberetal10-1,rebassa-mansergas16-1}, only about a dozen magnetic white dwarfs have been serendipitously identified \citep[e.g.][]{reimersetal99-1}. In these few detached magnetic white dwarf binaries, the main sequence star companions are close to Roche-lobe filling and the white dwarfs have effective temperatures below 10\,000 K \citep{parsonsetal21-1}. The only exception to this trend is the weakly magnetic white dwarf in the young post common envelope binary CC\,Cet \citep{Wilson21}. In the majority of the cold detached and strongly magnetic white dwarf binaries, the white dwarf rotation is synchronized with the orbital motion of the secondary star. The only clear exceptions are AR\,Sco, the first radio-pulsing white dwarf binary star \citep{marshetal16-1}, its recently discovered analogue, J191213.72−441045.1 \citep{Pelisoli23}, and 2MASS\,J0129+6715 which also shows some indications for non-synchronous rotation \citep{Hakala22}. 

Different again is the situation in close double white dwarfs. These systems must have evolved through two phases of mass transfer. Among the dozens of known detached close double white dwarf binaries, only one strongly magnetic white dwarf is known
\citep{kawkaetyal17-1, schreiberetal22-1}, This may  indicate that the fraction of systems with magnetic fields may be rather low or that detecting magnetic fields in double white dwarfs can be extremely challenging. 
Only very recently,
the first potentially weakly magnetic white dwarfs have been detected among semi-detached double white dwarfs, so-called AM\,CVn binaries \citep{maccaroneetal23-1}. 

Several ideas have been put forward to explain the origin of strongly magnetic white dwarfs. The three most popular scenarios that have been suggested in the last decades are (i) the fossil field scenario in which the magnetic field of the progenitor of the white dwarf is preserved during the white dwarf formation \citep[e.g.][]{angeletal81-1,braithwaite+spruit04-1,wickramisnghe+ferrario05-1}; ii) a dynamo generated during
common-envelope evolution in close binaries \citep{reghos+tout95-1,toutetal08-1,wikramasingheetal14-1} and (iii) coalescing double degenerate cores/objects \citep{garcia-berroretal12-1}.
However, all three scenarios face serious difficulties when compared to observations. 
The relative
numbers of strongly magnetic white dwarfs predicted by the fossil field scenario are far lower than the observed numbers if updated star formation rates and evolutionary time scales are taken into account \citep{kawka04-1}. The solution to this problem suggested by \citet{wickramisnghe+ferrario05-1}, who postulated the existence of a large number of main sequence stars slightly less magnetic than Ap and Bp stars, was refuted by spectropolarimetric surveys \citep{auriereetal07-1}. 


The common envelope dynamo scenario in its current form predicts relative numbers of magnetic systems far too large when compared to observations \citep{belloni+schreiber20-1}, and
the biggest weakness of the double degenerate merger scenario is that it cannot explain a large number of magnetic white dwarfs among CVs.
In addition, all three scenarios do not offer an explanation for the absence of young detached
magnetic white dwarf binaries \citep[e.g.][]{liebertetal05-1} and the late appearance of the magnetic fields in single white dwarfs \citep{bagnulo+landstreet21-1}.

Based on the idea originally put forward by \citet{isernetal17-1}, an alternative model to explain the incidence of magnetic fields in white dwarfs has been recently suggested by \citet{schreiberetal21-1}. 
This scenario has been shown to explain a large number of observations of magnetic white dwarfs in binaries: 
the increased occurrence rate of magnetic white dwarfs in CVs, the paucity of magnetic white dwarfs in the sample of observed double white dwarfs, the relatively large number of detached but close to Roche-lobe filling cold magnetic white dwarf plus M-dwarf binaries, 
the existence of radio-pulsating white dwarfs such as AR\,Sco \citep{schreiberetal21-1, schreiberetal21-2, schreiberetal22-1}, as well as the absence of high accretion rate polars in globular clusters \citep{bellonietal21-1}.  
One of the key predictions originally made by this scenario is that strongly magnetic crystallizing white dwarfs should rotate significantly faster than non-magnetic white dwarfs. 

However, \citet{ginzburgetal22-1} recently suggested that the convective turnover times in crystallizing white dwarfs are orders of magnitude longer than previously thought. If this is true, white dwarfs with spin periods of several hours or even days can 
generate magnetic fields of the order of a MG. If super-equipartition is assumed \citep{Augustson16}, even much stronger fields, such as those observed in many CVs, covering the range of 1-100\,MG can be produced. Crucial for the context of the present paper, this model predicts a relation between spin period and field strength. However, testing this hypothesis, i.e. faster rotation in magnetic than non-magnetic white dwarfs and/or a relation between field strength and rotation requires a representative sample of spin periods of magnetic white dwarfs. 

Magnetic white dwarfs can show photometric variability which allows for measuring their spin periods. 
This variability can have different origins. 
In convective atmospheres starspots can be generated by the magnetic field.   
As these regions are cooler and darker, starspots rotating into view, reduce the observed brightness. 
A strong magnetic field might however completely inhibit convection in the atmospheres, which makes the appearance of starspots 
unlikely \citep{Tremblayetal15}.
Alternatively, the Balmer lines can be split and shifted to the blue due to the presence of a strong magnetic field. The amount of this shift depends on the local field strength, which will change the spectral energy distribution locally (even if the flux remains unchanged), leading to light variation in observations in a single passband if the local field strength varies much over the stellar surface. This effect would predict larger variations (on average) in stars with stronger fields, and maybe larger amplitudes in hotter white dwarfs with stronger Balmer line blocking \citep{Hardyetal23}.
A third potential origin of photometric variability is that the polarized line opacities depend on the 
local field strength and on the angle a given region is looked at. The latter is probably a smaller effect, but might account 
for variations of the order of one per cent in some cases.
Finally, magnetic fields can cause metals to be distributed nonhomogenously on the white dwarf surface which can cause photometric variations as well \citep{dupuisetal00-1}.
Independent of the exact mechanism producing the variability, photometric variability of magnetic white dwarfs allows for measuring their rotation rates. 

The volume-limited sample of
white dwarfs within 20\,pc contains 33 magnetic white dwarfs \citep{bagnulo+landstreet21-1}.  
This sample is ideal to study by measuring the rotation periods of a representative sample of magnetic white dwarfs thereby potentially constraining scenarios for the origin of magnetic fields. 
We found that 27 of these magnetic white dwarfs have been observed with {\it TESS} \citep[Transiting Exoplanet Survey Satellite,][]{Ricker15}. The resulting light curves show statistically significant and constant periodic signals (the same period in all {\it TESS} sectors) in only five cases. Given this small sample size, we included all known magnetic white dwarfs within 40\,pc and identified three more periods.
In addition to analyzing the {\it TESS} light curves, we followed up two additional targets (one of them part of the 20\,pc sample) with SPECULOOS \citep[Search for habitable Planets EClipsing ULtra-cOOl Star,][]{Delrez18, Jehin18}. 
We compared our period measurements to those non-magnetic and magnetic white dwarfs with previously measured spin periods, and finally, we discuss possible implications for the origin of magnetic fields in white dwarfs. 


\section{Observations}

In this work, we combine photometric data of magnetic white dwarfs from {\it TESS} with light curves obtained using the SPECULOOS instrument. In what follows we briefly describe the data acquisition for both cases as well as the procedure we used for determining the rotational period.

\subsection{{\it TESS} } 

For all magnetic white dwarfs within 20\,pc listed in \citet{bagnulo+landstreet21-1}  we searched for {\it TESS} light curves. Of the 33 targets on the list, we found {\it TESS} observations for 27 magnetic white dwarfs which are listed in Table\,\ref{tab:secotrs20} with their corresponding sectors.  We also took a careful look at the new magnetic white dwarfs identified in the 40\,pc sample \citep{OBrien23, bagnulo+landstreet22-1}
{\it TESS} data are available for 23 of the 30 new magnetic white dwarfs \citep[][their Table
4]{OBrien23} and for eight white dwarfs from \citet{bagnulo+landstreet22-1}, all listed in Table\,\ref{tab:secotrs40}.  

The {\it TESS} light curves were obtained from the Mikulski Archive for Space Telescopes (MAST\footnote{https://mast.stsci.edu}) web service. We extracted the Presearch Data Conditioned Simple Aperture Photometry (PDCSAP) which removes trends caused by the spacecraft, and removed all data points with a nonzero quality flag and all NaN values in each sector. 

Contaminating flux from unexpected sources which occurs due to a combination of pixel size and flux integration is a known issue in {\it TESS} light curves. We therefore performed a test to identify possible contaminating flux in the light curves using the  flux contamination tool\footnote{
https://www.jessicastasik.com/flux-contamination-tool
} \citep[{\sc FluxCT,}][]{Schonhut23}. 
We note that this tool is based on {\it Gaia} G-band magnitudes of the objects in each pixel, i.e. the tool is using a band-pass different from {\it TESS}. Therefore, the estimated levels of contamination may not be entirely accurate. How much the real values deviate from the estimates depends on the colors of the source and contaminants. However, given that the two bands overlap, our estimates should not differ from the real contamination level by more than a few per cent in most cases. 

We emphasize that a critical examination of the data of each target is fundamental to avoid wrong conclusions being drawn. To that end, we slightly modified 
the code that provides contamination levels for {\it TESS} targets. 
The original version only offers the contamination level and the {\it Gaia} G-magnitude of the target for the first observed sector. More insight can be gained by providing the contamination level and G-magnitude of the target for each sector.

We analyzed each sector of each target with the least-squares spectral method based on the classical Lomb-Scargle periodogram \citep{Lomb76,Scargle82} to obtain the main period of the photometric {\it TESS } data. For each star, we started with the least contaminated sector to make sure the signal we are picking up is coming from the white dwarf and then requested the detected period to show up with a consistent amplitude in all other sectors. We searched for periods in frequency space up to the Nyquist frequency.  

To make sure the signal is significant, we also performed a false alarm probability (FAP) test. We formally requested 
this false alarm probability to be below 5 per cent. For all the periods we detected we found the FAP to be less than $10^{-6}$. 
The uncertainties of the periods were computed using the curfit routine from \citet{Bevington69}, which is a Levenberg-Marquardt non-linear least-squares fitting procedure.

In case a given white dwarf light curve passed all the above tests 
we finally inspected adjacent {\it TESS} pixels to check whether nearby bright stars could have contaminated the white dwarf light curve. 
The {\sc FluxCT} mentioned above only provides information on stars located within the same {\it TESS} pixel. 
We therefore used the {\sc lightkurve} tool \citep{Lightkurve18} for this exercise. 
If a bright source was found we downloaded its {\it TESS} light curve (in case available) and ran a period search. If the same period was found as for the white dwarf, we used the amplitude of the variation to decide if the photometric variability is indeed coming from the white dwarf.  

As a first example that illustrates the importance of a careful analysis of {\it TESS} data, 
we show in the appendix (Figure \ref{fig:false}) the light curve we obtained for WD\,2150+591. 
A very strong signal is clearly present in the data with a period of 116.38\,hr. 
However, the amplitude largely exceeds those found for other white dwarfs and the light curve resembles that of an eclipsing binary. The up-dated {\sc FluxCT} provided the G-magnitude for the target of each sector which revealed that the G-magnitude of the target in the first sector was clearly different to that of WD\,2150+591. In other words, the detected star in the first sector was a nearby eclipsing binary instead of the white dwarf we aimed to analyze, and we therefore eliminated this white dwarf from our sample of systems with measured period. 

\begin{figure*}
        \centering
    \includegraphics[width=1.9\columnwidth]{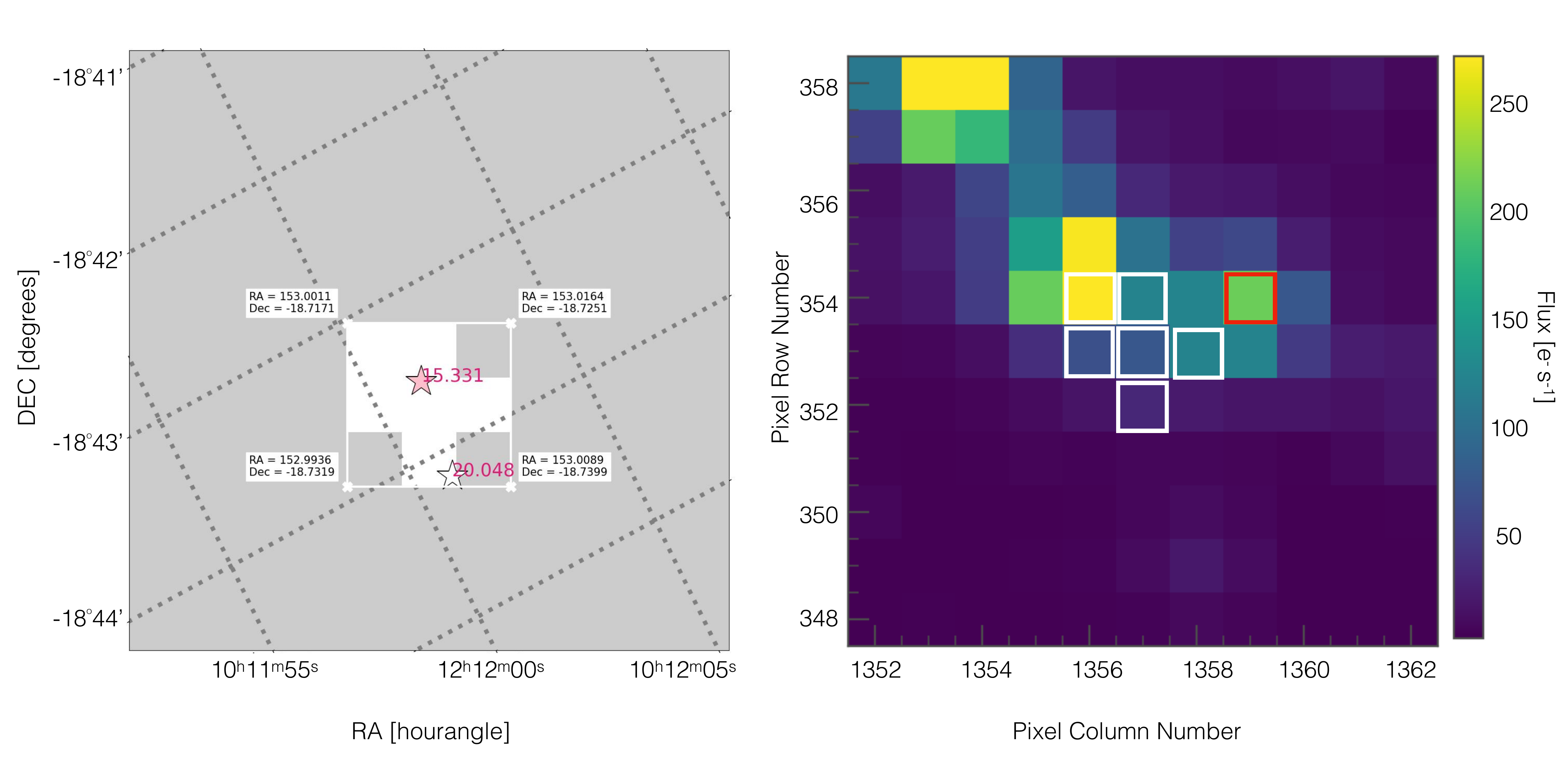}
    \caption{ 
    The pixel view obtained with the two different tools used to study the contamination from nearby sources to the white dwarf observed with {\it TESS}. {\it Left:}  The {\sc FluxCT} provides a white area corresponding to the pixels used to calculate the contamination. The position of WD\,1009-184 is indicated by a pink star while that of the only contaminating source found by the tool is shown as a white star. The numbers right next to each star indicate their {\it Gaia} G-magnitudes. {\it Right:} The image obtained with the {\sc lightkurve tool}. Pixels with a white edge were used to extract the {\it TESS} light curve of WD\,1009-184. The variation found in the light curve of this system, however, results from contamination of these pixels from a bright star in the pixel with the red border. The star located in this pixel shows cleaner variations with the same period and a larger amplitude. This type of contamination cannot be identified if {\it TESS} data is only analyzed with the {\sc FluxCT}.   However, {\sc tess\_localize} confirms that the origin of the variation corresponds to target Gaia\,DR3\,5669427508702256896 located at the pixel  marked with the red border.}
    \label{fig:pixeles}
\end{figure*}

Taking a detailed look at adjacent pixels with the {\sc lightkurve tool} turned out to be important as well.  
In one case (WD\,1009-184), we indeed found that the period measured from the white dwarf light curve most likely corresponds to that of a nearby bright star (Figure\,\ref{fig:pixeles}) and we excluded the white dwarf from our sample of stars with reliable periods. \citet{bagnulo+landstreet19-1} report variations on the magnetic field strength based on two measurements for this white dwarf but more measurements are needed to constrain the spin period. 

 Up to this point, we have identified a total of 18 white dwarfs exhibiting significant variability: 11 within 20\,pc and 7 within 40\,pc. For a final decisive test, we 
analyzed all 18 white dwarfs with the new tool {\sc tess\_localize}\footnote{github.com/Higgins00/TESS-Localize} \citep{Higgins23}. This tool is especially designed to localize the source most likely responsible 
for observed variations in each {\it TESS} pixel. {\sc tess\_localize} delivers the optimized column and row coordinates corresponding to the most probable location of the observed variability. The algorithm complements {\it TESS} using queries to the {\it Gaia} Archive\footnote{https://gea.esac.esa.int/archive/} \citep{gaiaDistance21} for star locations and offers metrics such as p-values and relative likelihoods to facilitate interpretation of the fit outcomes. Prior to the fitting process, the {\sc tess\_localize} tool offers the option to discern prevalent trends among pixels located outside the designated aperture. This task is accomplished through principal component analysis (PCA). The resulting PCA components can be effectively applied to and subtracted from the light curves extracted by {\sc tess\_localize}. Nonetheless, it is imperative to ensure that these PCA trends do not represent the signals targeted for localization; otherwise, the signals may be inadvertently removed from the data. 
It is important to note that {\sc tess\_localize} should provide a substantial detection, meaning that the 'height' parameter in the fit is significantly different from zero given its uncertainty, ensuring it is not a false positive detection.

Eighteen targets were initially considered but 10 were subsequently eliminated from the sample due to {\sc tess\_localize} results showing that the signals observed in the lightcurves did not originate from the white dwarf targets. These eliminated targets are as follows: WD\,0810-353, WD\,0816-310, WD\,1036-204, WD\,1829-547, WD\,1900+705 and WD\,2153-512 from the 20\,pc sample and WD\,0232+525, WD\,1008-242, WD\,J091808.59-443724.25, and WD\,J094240.23-463717.68 from the 40\,pc sample. 
For all the white dwarfs discarded with {\sc tess\_localize}, the origin of the detected variation was well located on the field with the exception of 
WD\,1008-242 where we were unable to identify the source of the measured variability.

\begin{figure*}
    \centering
    \includegraphics[width=1.9\columnwidth]{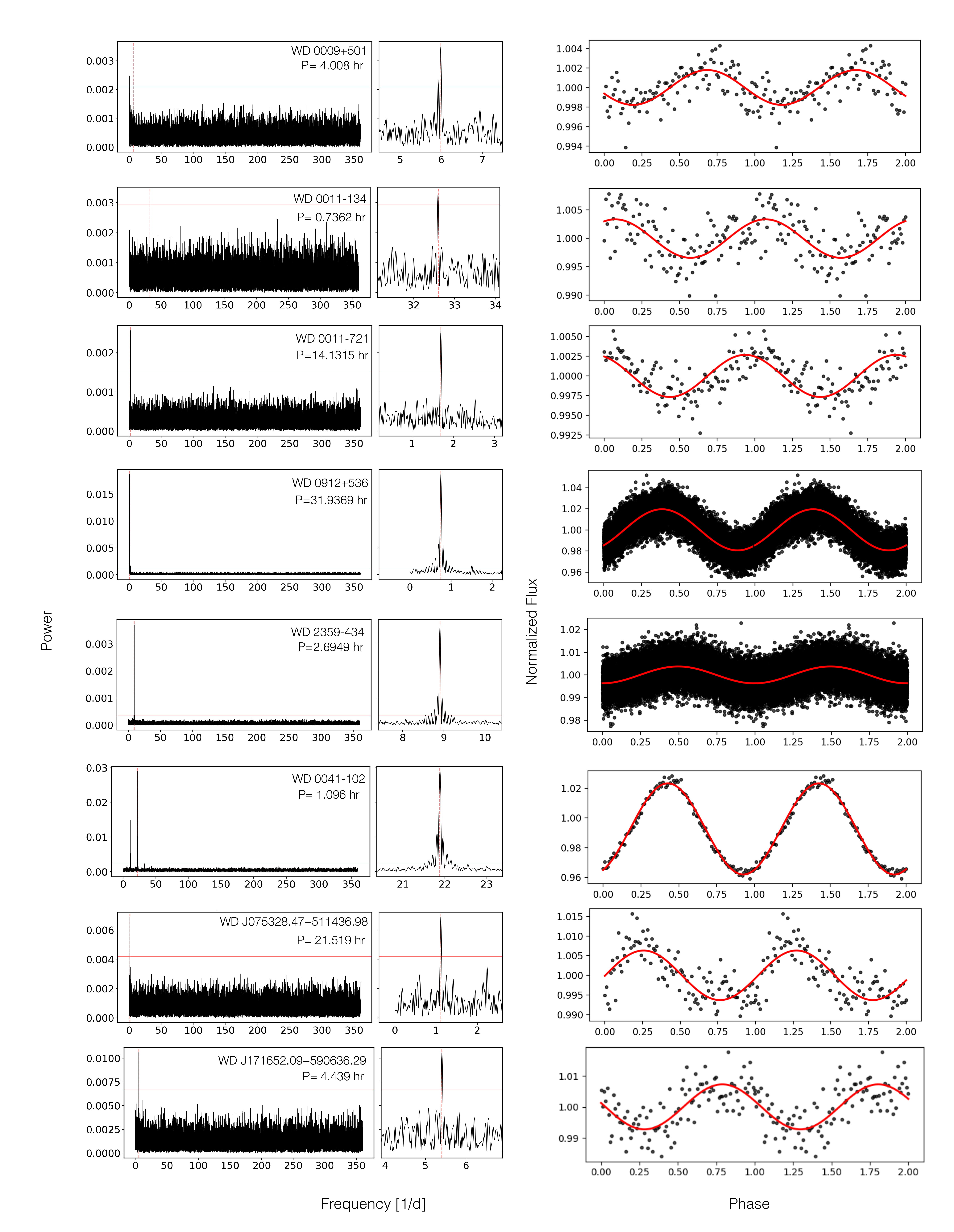}
    \caption{Periodograms  (left: full Nyquist range; middle: zoomed in to the highest peak) and phase-folded light curves (right) of  eight white dwarfs with a {\it TESS} light curve that shows statistically significant variations with the same period in all available sectors.  The horizontal red line illustrates the false alarm probability level (FAP level) for a probability of $10^{-3}$ per cent. 
    In the case of WD\,0009+501, WD\,0011-134, WD\,0011-721, WD\,0041-102, WD\,J075328.47-511436.96 and WD\,J171652.09-590636.29 we binned the phase-folded light curve (using 100 bins) as the amplitude of the variation is too small to be spotted in the un-binned light curve.   For WD\,0009+501 and WD\,0041-102 we know from 
    previous observations that the true period is twice the period we detect in the {\it TESS} data \citep{valyavinetal05-1, achilleosetal92-1}. We use the longer (and correct) periods in the distributions discussed in this paper. 
    The {\it TESS} light curves folded over the rotation period can be found in the appendix. For the other seven objects, we interpret the observed periods as reflecting the rotation periods of the white dwarfs. The measured periods range from a few hours to several days. 
    }
        \label{fig:TESS_LC}
\end{figure*}

 Considering all tests listed above, the final sample of white dwarf periods measured from {\it TESS} contains eight white dwarfs, five with distance less than 20\,pc and three within the 40\,pc. For these systems the average G-magnitude in the {\it TESS} sectors is in agreement with the one reported in literature for the corresponding white dwarf and contamination from nearby sources can be excluded as the source of the identified variations. 
The contamination level and the exposure times of each {\it TESS} sector  of the confirmed white dwarfs are shown in 
Table\,\ref{tab:secotrs20} and \ref{tab:secotrs40}.
%
%
%
The measured  periods,  G magnitudes and normalized amplitudes obtained from {\it TESS} light curves as well as the frequencies and PCA entries for the {\sc tess\_localize} tool and the best fits results (p-values and likelihood) 
for the  eight targets with clear, significant and consistent variations (that can be assumed to reflect the rotation period) 
are listed in Table\,\ref{tab:Localize}. The periodograms and the phase-folded light-curves (which include all the sectors mentioned in Table\,\ref{tab:secotrs20} and \ref{tab:secotrs40}) for each of the  eight {\it TESS} targets are shown in Figure\,\ref{fig:TESS_LC}.

\begin{table*}
    \caption{ Rotation periods measured from {\it TESS} and SPECULOOS together with their normalized amplitudes. We provide two period approximations for WD\,2138-332, i.e. the period corresponding to highest peak in the periodogram which represents the best fit and the range of possible rotational periods given the data currently available. The extra line for WD\,0009+501 and WD\,0041-102 provides the rotational period measured through independent  observations by 
    Bagnulo, Landstreet $\&$ Valyavin (in preparation) and \citet{achilleosetal92-1} which correspond to twice the period measured with {\it TESS} ( see section \ref{sec:second_peak} for more details). We provide the entries for the {\sc tess\_localize} tool; TIC name, the frequency measured from the {\it TESS} light curves, and the number of signals removed (PCA) to obtain the best signal to noise for the signal fitting. Finally, we present the resulting p-values and relative likelihoods for the best fit indicating that the white dwarf is the variable source.  }
    \centering
    \resizebox{17.5cm}{!}{\begin{tabular}{llllllllll}
    \hline
         Name & TIC name & Gaia DR3 & G & Period& Frequency & PCA & p-value & Likelihood  & Normalize \\
           &  &   & [mag] & [hr] &[$\mu$Hz] &    &   & [\%] & amplitude\\
        \hline
     {\it TESS} &&&&&&&&&\\
     \hline
         WD\,0009+501 & TIC\,201892746 &    395234439752169344 &     14.23 &  $4.0086 \pm 0.0001 $ & 69.2954592071 & 3& 0.597 & 0.99 & 0.0034$\pm$ 0.0003 \\
         &&&&8.016 $\pm$ 0.083 &34.69620007& 0 & 0.038& 0.99& 0.0042 $\pm$ 0.0002 \\
         WD\,0011-134  & TIC\,289712694 &   2418116963320446720 &    15.75 &  $0.736 \pm 0.007$ &  377.415458937& 0& 0.536 & 1.0  & 0.0033$\pm$ 0.0004   \\
         WD\,0011-721 & TIC\,328029653	&  4689789625044431616  &    15.03 &   $14.13 \pm 0.38$ & 19.6566378500 & 0& 0.127 & 0.99 & 0.0038$\pm$ 0.0003 \\
         WD\,0912+536 & TIC\,251080865  &   1022780838739029120 &    13.78 &   $31.93 \pm 0.13$  & 8.69795145847 & 0& 0.667 & 1.0 &  0.0189$\pm$0.0001   \\
         WD\,2359-434 & TIC\,321979116	&   4994877094997259264 &    12.89 &    2.694 $\pm 0.002$ & 103.109791305 & 0& 0.413 & 1.0   & 0.0037 $\pm$0.0001  \\
         WD\,0041-102 & TIC\,3888273	& 2377863773908424448  & 14.5 & 1.0967 $\pm$ 0.0007 & 253.44687753 & 0 & 0.542 &1.0 & 0.0290 $\pm$ 0.0003 \\
            &&&& 2.19$\pm$1.09  & 126.723438767 & 0 & 0.379 & 1.0 & 0.0327 $\pm$ 0.0003\\
         WD\,J075328.47–511436.98 & TIC\,269071459 & 5513896164414899456 & 15.6  & 21.51$\pm$ 0.01 & 12.9084891387 & 2 & 0.369 & 1.0 & 0.0062$\pm$0.0005\\
         WD\,J171652.09–590636.29 & TIC\,380174982 & 5915797694789556096 & 15.62 & 4.4398$\pm$0.0003  & 62.5653808229 & 0 & 0.297	& 0.96 & 0.0080$\pm$ 0.0001 \\
    \hline
    SPECULOOS &&&&&&&&&\\
    \hline
        LSPM\,J0107+2650 & - & 306779618349361920 & 18.88 & 4.83 $\pm$ 0.29 & -&- &-&-&   0.11$\pm$ 0.01  \\
        WD\,2138-332 & TIC\,204440456 & 6592315723192176896 & 14.44 & 6.19 $\pm$ 0.05 & -&- &-&-& 0.008 $\pm$ 0.001 \\ 
        &&&&4.0-12.0&&&&&\\
    \hline
    \end{tabular}}

    \label{tab:Localize}
\end{table*}

The periodicity in the light curves of WD\,0009+501, WD\,0011-134, WD\,0011-721, WD\,J075328.47−511436.98 and WD\,J171652.09−590636.29 
is difficult/impossible to spot in the phase folded light curves. 
 With the aim to provide a better visualization of the five mentioned light curves, we therefore 
calculated the mean flux for $100$ bins. The corresponding folded light curves
are shown in 
Fig.\,\ref{fig:TESS_LC}. 
As mentioned earlier, the false alarm probability for the highest peak in each periodogram is below $10^{-6}$.  We illustrate the FAP level corresponding to a probability of $10^{-3}$
in all periodograms in 
Fig.\,\ref{fig:TESS_LC}. We note that in two cases, WD\,0009+501 and WD\,0041-102, a second highly significant peak appears in the periodograms. Previous observations show that these second highest peaks correspond to the rotation period of these two white dwarfs 
\citep{valyavinetal05-1, achilleosetal92-1}. 
The light curves folded over the true rotation period are shown in Figure\,\ref{fig:TESS_LC_r} in the appendix. 


\subsection{SPECULOOS}

For two white dwarfs, WD\,2138-332 (part of the 20\,pc sample) and LSPM\,J0107+2650 (a more distant magnetic DZ white dwarf), we performed observations with the SPECULOOS Southern Observatory (SSO). SPECULOOS is composed of four telescopes with primary mirrors of 1.0\,m diameter, run by the European Southern Observatory (ESO) and located at Paranal, Chile. 
We used the SPECULOOS/Y4KCam CCD with the {\it SDSS\, g'} filter. This filter covers the range where the Balmer lines $\beta  $ and $ \, \gamma$, and some of the strongest Helium lines, can be found when observing white dwarfs. The dates, exposure times, and the name of the telescope that was used are listed in Table\,\ref{tab:SPECULOOS}.

\begin{figure*}
    \centering    
    \includegraphics[width=2.0\columnwidth]{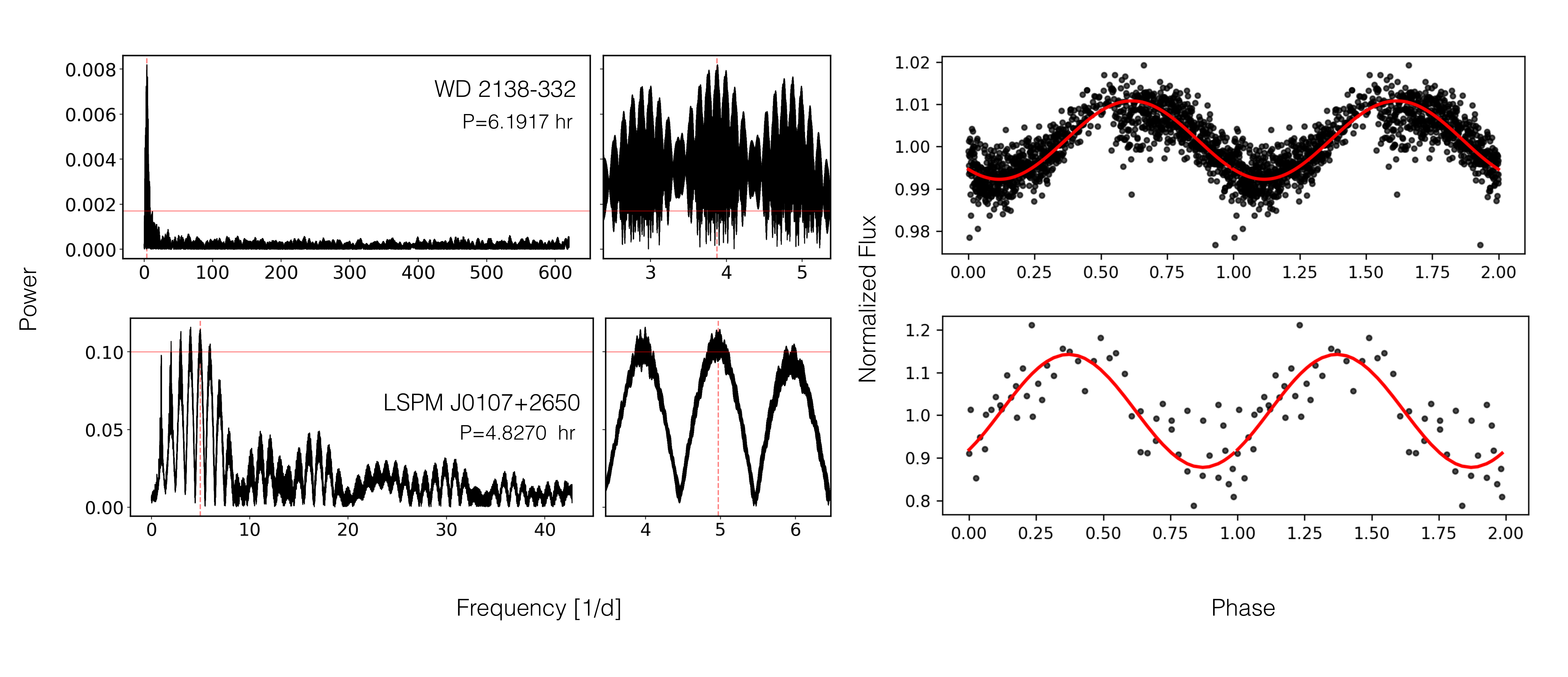}
    \caption{Periodograms  (left: full Nyquist range; middle: zoomed in to the highest peak) and phase-folded light curves  (right) of the two white dwarfs observed with SPECULOOS. The obtained periods are less certain than those measured from {\it TESS} light curves.  The horizontal grey line illustrates the false alarm probability level (FAP level) for a probability of $10^{-6}$ per cent. 
    For WD\,2138-332 we can estimate only a range of periods because several periods (peaks in the periodogram) provide reasonable fits to the data. However, both targets clearly have periods of the order of a few hours.  } 
    \label{fig:SPECULOOS_LC}
\end{figure*}

Data reduction was performed using {\sc prose} \citep{Garcia22}, a Python framework to build a modular and maintainable image processing pipeline. Dark, bias and flat field corrections were executed to all the images, followed by an automated line-up in preparation for the aperture photometry. We then analyzed the resulting photometry with the least-squares spectral method used for {\it TESS} targets and found that both targets showed clear variation associated with the rotation of the white dwarf. For WD\,2138-332, the periodogram shows several aliases, and the light curve fit is reasonable with a number of different periods between 4 and 12 hr, with 6.19 hr corresponding to the highest peak in the periodogram providing the best fit. However, with the data currently available, we can only constrain the period to be between 4-12 hr. 
Spectral type, effective temperature, mass and magnetic field strengths \citep{hollandsetal17-1,bagnulo+landstreet19-1} together with the rotational period measured with SPECULOOS are listed in Table\,\ref{tab:Localize}, while the periodograms and the phase-folded light curves are shown in Figure\,\ref{fig:SPECULOOS_LC}.

\section{The rotation periods of magnetic white dwarfs} 

We have measured periodic photometric variations for ten magnetic white dwarfs,   five of which are within a distance of 20\,pc. 
In addition,  eight of these periods have been obtained from {\it TESS} light curves which cover at least one month of continuous observations and are therefore not biased towards short periods, in contrast to those determined from short (a few days) observing runs using ground-based telescopes. 
%
%
In what follows we compare our findings with periods of magnetic and non-magnetic white dwarfs provided in the literature. 

Most rotation periods of magnetic white dwarfs result from photometric measurements with ground-based telescopes such as the two periods we determined with SPECULOOS. Additional periods have been determined through variations in the polarization degree. Given the fact that it is virtually impossible to use ground-based telescopes to reach the same cadence and baseline as the {\it Kepler} surveys or {\it TESS}, the full sample of rotation periods of magnetic white dwarfs is likely biased towards shorter periods when compared to that of non-magnetic white dwarfs. 

We used two different approaches towards a more representative samples of magnetic white dwarf spin periods.  First, we only used periods measured from {\it TESS} light curves, except for WD\,0009+501, whose period measured with {\it TESS} is half of the real period measured with spectropolarimetry. The cadence and baseline of {\it TESS} is much longer than the periods measured for non-magnetic white dwarfs. Therefore, the  eight  periods we measured from {\it TESS} (Table\,\ref{tab:Localize}) should not be biased towards shorter periods, and thanks to the very short exposure time in some of the sectors (20\,sec), the chances to miss an extremely short spin period are low. The obvious disadvantage of this sample is its small size. 

Because of the small size of the {\it TESS} sample, we also establish a volume-limited sample consisting of spin periods of magnetic white dwarfs within $20$\,pc. This sample combines  five period measurements from {\it TESS}, the rough period estimate we obtained for WD\,2138-332, and four periods of magnetic white dwarfs within $20$\,pc from the literature ( Table\,\ref{tab:brink}). 
This results in a sample of  ten magnetic white dwarfs with measured rotation period within 20\,pc which 
represents $30$\,per cent of the 20\,pc sample of magnetic white dwarfs. The disadvantage of this sample is that it is incomplete and might again be biased towards short periods.

\subsection{Comparison with previously identified periods of magnetic white dwarfs}

To compare our results with previously measured rotation periods of magnetic white dwarfs, we established a list of robustly measured rotational periods from the literature, i.e. in what follows we ignore uncertain period estimates or measurements that provided a range of possible periods. Our final sample of magnetic white dwarfs with robustly measured periods from the literature is given in Table\,\ref{tab:brink}. 
It contains seven white dwarfs from \citet{brinkworthetal13-1}, who
presented observations of 30 isolated magnetic white dwarfs performed with the 
Jacobus Kapteyn Telescope and compiled a list of previously published periods. 
We further complemented our literature search by going through the periods listed in \citet{kawkaetal07-1} and \citet{ferrarioetal15-1} -- ignoring rather rough estimates -- and by adding more recent measurements. In addition, we list three yet unpublished periods measured by Landstreet and Bagnulo through polarimetry. The observations that allowed the determination of these periods will be published elsewhere.
 This compilation of literature encompasses a  total of 42 white dwarfs with reliably measured spin periods. 
Five of these 42
are in our {\it TESS} sample. 
Six of these 42 magnetic white dwarfs are very likely the product of the merger of two degenerate stars, as they are very massive white dwarfs ($>1.25$\,\Msun) with rotation periods below $0.38$\,hr. An atypical result of a merger is SDSS\,J125230.93-023417.72 with low mass 
\citep[$0.58\,\Msun$, ][]{Reding23}.
Occasionally we eliminate these likely merger products from the sample in order to constrain other possible mechanisms for generating the white dwarf magnetic fields. 

We start the comparison of our samples with previously published spin periods with WD\,0912+536. 
This white dwarf was one of the first known magnetic white dwarfs and was found to show periodic variations in circularly polarized light with a period of $=32.16$\,hr \citep{angel+landstreet71-1}.
The most obvious explanation for periodic variations in polarized light is the rotation of the white dwarf in combination with a magnetic field that is not symmetrical about the spin axis. The period we determined from the {\it{TESS}} light curve (31.93\,hr) is very close to the value measured by \citet{angel+landstreet71-1}. This agreement confirms that we can indeed assume periodic light curve variations to reflect the rotational periods. Other examples for spin periods measured with {\it TESS} and through polarimetry are WD\,0011-134 and 
WD\,2359-434, in both cases the periods also agree (compare Table\,\ref{tab:Localize} with Table\,\ref{tab:brink}).
An interesting example is WD\,0009+501. For this star \citet{Valyavin05} finds the magnetic field to be variable with a period of eight hours and \citet{Valeev15} found photometric variability confirming this period but mentioned that the light curve showed two maxima per period. The {\it TESS} light curve confirms this finding but without the knowledge of the field strength variability, we would have interpreted the stronger periodic signal as the rotation period ($4$\,hr) which is in fact half the real rotation period.  The {\it{TESS}} light curve folded over the true rotation period is shown in Fig.\,\ref{fig:TESS_LC_r} in the appendix. 

 Also the highest peak in the periodogram of WD\,0041--102 corresponds to half the rotational period previously measured. If we fold the light curve over the period derived by \citet{achilleosetal92-1}, we find two humps of different amplitudes in agreement with their results (see Fig.\,\ref{fig:TESS_LC_r} in the appendix).
\citet{achilleosetal92-1} showed that this variation is produced by a magnetic field that is dipolar and orientated at about 90 degrees to the spin axis. Thus, as the star spins, the abundance pattern on the surface seen from Earth varies. The strength of spectral lines, the flux they block directly, and the line blocking blueward of the Balmer jump, as seen from Earth, also change as the star rotates. 
This most likely produces the very strong light variability of this star. For an alternative interpretation for white dwarfs displaying two minima in their light curves 
see 
\citet{Farihi23}.

It is worth pointing out that a similar issue
as identified for WD\,0009+501 and 
WD\,0041-102 could affect virtually all our other targets (i.e. we could be off by a factor of 2 for most objects). However, even if this is the case the conclusions
presented in this paper would not change.


To evaluate whether the {\it TESS} and the 20\,pc samples we defined are systematically different to the periods of magnetic white dwarfs in the literature we performed a Kolmogorov-Smirnov (KS) test and show the cumulative distributions in Figure\,\ref{fig:CDF}. Assuming the usual 2-sigma significance criterion (p-values below 0.05) to reject the Null-hypothesis (both distributions come from the same parent distribution), both our samples do not show significant indications for being different to that of magnetic white dwarfs we found in the literature (Table\,\ref{tab:KS}).  More precisely, the observed differences have an unacceptable high probability to be caused by chance (exceeding $30$ per cent). We also compare the average and median values of our samples and those of previously identified periods of magnetic white dwarfs and find comparable values (Table\,\ref{tab:KS}). For this exercise, we also excluded magnetic white dwarfs that are very likely the product of a merger (very fast-spinning, i.e. $P<0.38$\,hr and very massive, i.e. $M>1.25$\,\Msun, white dwarfs) in agreement with the predictions made by \citet{schwab21-1}. These white dwarfs are WD\,0316-849, WD\,1859+148, WD\,2254+076, WD\,2209+113, Gaia\,DR3\,4479342339285057408, and ZTF\,J1901+1458. While the median values in our samples are slightly longer than those of previously identified periods (even if we exclude mergers), the average periods are somewhat shorter. 
We conclude that the more representative samples defined here do not show significant differences to previously measured periods of magnetic white dwarfs.



\subsection{Comparison with non-magnetic white dwarfs} 

\citet{hermesetal17-1} and \citet{kawaler15-1} used data from the {\it{Kepler}} space telescope to measure the spin periods of pulsating non-magnetic white dwarfs
and found that the mean spin period of isolated non-magnetic white dwarfs with masses in the range of $0.51-0.72\,\mathrm{M_{\odot}}$ is $\simeq\,35$\,hr with a standard deviation of 28\,hr. In other words, the majority of non-magnetic white dwarfs have spin periods between 0.5-3 days. In the context of the crystallization and rotation-driven dynamo, and the origin of magnetic fields in white dwarfs in general, it is of fundamental importance to compare spin periods of non-magnetic and magnetic white dwarfs. 

In order to perform the comparison of spin periods of non-magnetic and magnetic white dwarfs, we need to consider possible observational biases. The rotation periods of non-magnetic white dwarfs have been measured through asteroseismology. The 36 non-magnetic white dwarfs with asteroseismological measurements of their rotation period are listed for completeness in Table\,\ref{tab:nomag}. They cover the mass range from $0.45-0.88\,\mathrm{M_{\odot}}$, which covers the peak of the mass distribution of isolated white dwarfs. Pulsating white dwarfs are located in the instability strip, and therefore cover a rather small range of temperatures (between $10\,000$ and $14\,000$\,K, depending slightly on the surface gravity). 
However, this bias with respect to temperature (and therefore age) is unlikely to affect the distribution of spin periods given the absence of an efficient braking mechanism such as, for example, magnetic braking. 

 On the other hand, given that magnetic white dwarfs  within 20\,pc are on average older \citep{bagnulo+landstreet22-1}, we need to consider that they may have experienced spin-down through the same process assumed for pulsars, that is, magnetic dipole radiation. 
This means that the initial spin periods of magnetic white dwarfs could have been shorter than the ones we measure today. To explore this possibility, we calculated the initial spin period of the white dwarf using Equation 2 in section 3.3 from \citet{williamsetal22-1}, which requires the mass, radius, and the measured spin period (Table\,\ref{tab:OBS}). We then integrated the equation over the cooling age and found that for our targets, the initial spin period was not significantly different (less than 5 per cent shorter) from the one currently measured, except for WD 0011-134, which 
has a short period of 0.74\,hr and is more than 4\,Gyr old. This white dwarf might have been fast spinning initially. 
However, there is strong evidence for the late appearance of the magnetic field in white dwarfs \citep{schreiberetal21-1,bagnulo+landstreet21-1}, which means that white dwarfs we observe today as magnetic white dwarfs, very likely did not emit magnetic dipole radiation throughout their entire cooling age but only since the magnetic field emerged. Therefore, reconstructing initial spin periods remains impossible as long as we do not understand the mechanisms responsible for the magnetic fields in white dwarfs. 

We are aware of the fact that the rotation rates from photometric variations and those inferred from asteroseismology may be measuring slightly different quantities, i.e., surface and globally averaged rotation rates, respectively \citep{oliveraetal22-1}. These quantities might be slightly different but we expect 
them to be in general very similar. 
 This assumption seems to be justified as \citet{hermesetal17-2} could show 
for the helium-atmosphere white dwarf 
PG\,0112$+$104 that the rotation period derived from the light curve ($10.17$\,hr) was 
consistent with the rotational splittings from pulsations which indicate a period of $\sim10$\,hr.
We therefore decided to compare the currently measured spin periods of magnetic white dwarfs with that of non-magnetic white dwarfs. Neither the fact that the latter are systematically younger nor that the periods derived from light curves and pulsations measure slightly different quantities affects the conclusions we draw in this paper.

\begin{figure}
	\includegraphics[width=1\columnwidth]{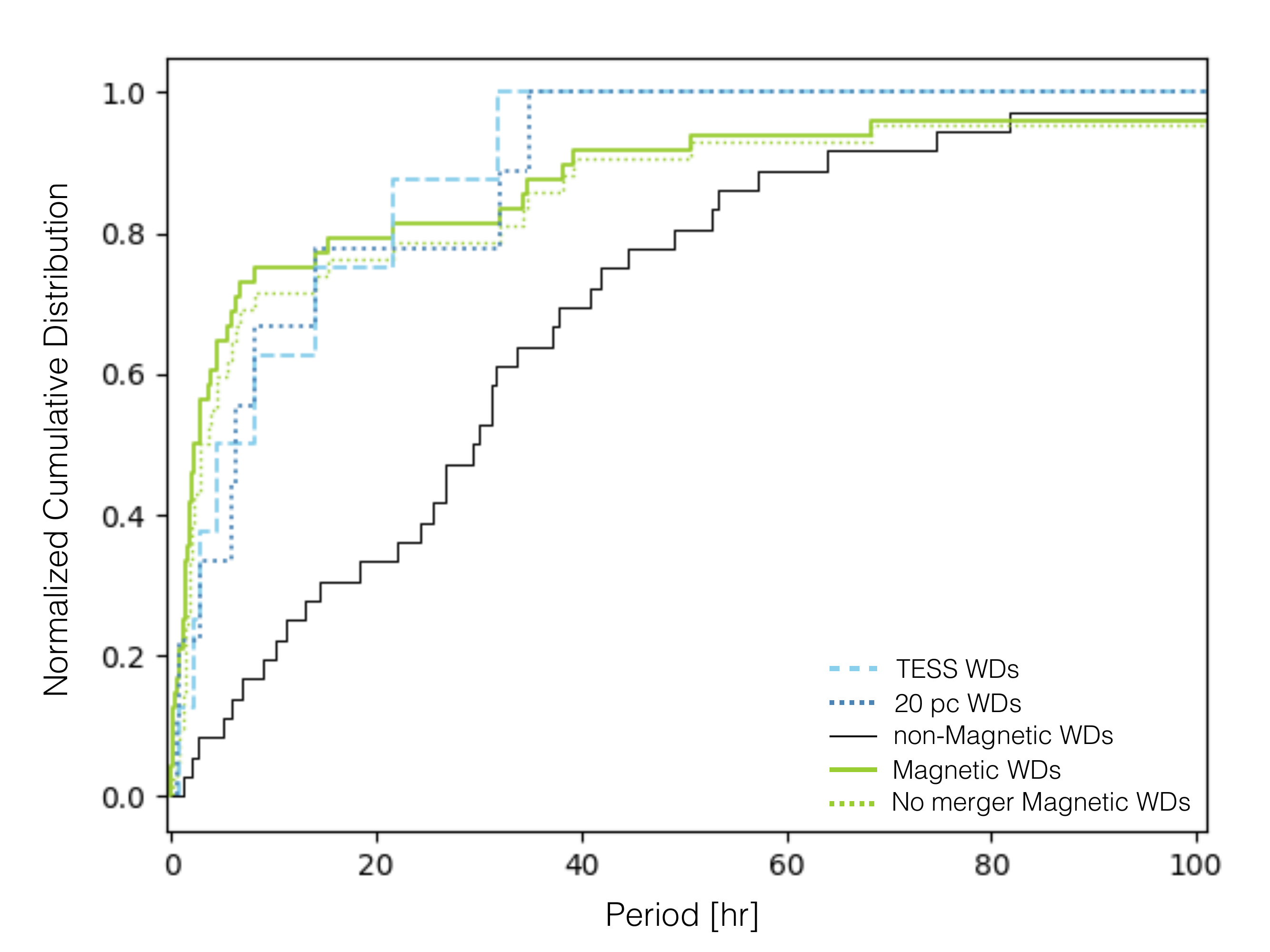}
    \caption{Cumulative period distributions of magnetic and non-magnetic white dwarfs. If we consider all rotation periods measured for magnetic white dwarfs (solid green line)  and the magnetic white dwarf after removing white dwarfs that most likely result from mergers (dotted green line), they seem to rotate significantly faster than non-magnetic white dwarfs (solid black line). A KS-test between the full magnetic and non-magnetic sample gives a probability of only  3.76$\times 10^{-7}$ for the observed difference to be caused by chance.  
    However, this is clearly caused by an observational bias. If only periods measured with {\it TESS } are considered (dashed light-blue line), a KS-test provides  weak indications for the period distribution being different to those of non-magnetic white dwarfs {(p-value $= 0.027$)}. The incomplete sample of magnetic white dwarfs within 20\,pc (dotted dark-blue line) also contains slightly shorter periods compared to non-magnetic white dwarfs.
    }
    \label{fig:CDF}
\end{figure}

The corresponding cumulative distributions are shown in Figure\,\ref{fig:CDF}. We used the Kolmogorov-Smirnov test to statistically evaluate possible differences between the samples. The results of this test are listed in Table\,\ref{tab:KS}. The full sample of measured periods for magnetic white dwarfs is clearly different to the rotation rates of non-magnetic white dwarfs with a high significance  ($p\simeq 3.76 \times 10^{-7}$). As the periods of the magnetic sample are potentially biased towards short periods (mostly coming from ground-based observations), this result does not allow to derive strong conclusions.

{\it TESS} provides continuous light curves for long enough time spans that in principal allow to measure periods of several days and therefore this sample is not biased against periods as long as those measured for non-magnetic white dwarfs. The range of periods we measured with {\it{TESS}} still appears to be systematically shorter than that of non-magnetic white dwarfs. While the latter cover periods from $0.33-2.34$\,days, the {\it TESS} periods we present here range from $0.01$ to $0.88$\,days. The KS tests comparing the periods of non-magnetic white dwarfs with those we measured here confirms these differences (the $p$-value is reaching $2\sigma$ significance). 
The median and average values further confirm the impression that magnetic white dwarfs rotate on average faster than non-magnetic white dwarfs (Table\,\ref{tab:KS}).


Our measurements, however, do not allow us to draw strong conclusions. 
First, our samples are simply rather small.
To provide further evidence for our findings, measuring the rotation periods of all magnetic white dwarfs in the 20\,pc sample, e.g through polarimetry, therefore represents a necessary next step which, unfortunately, requires relatively large amounts of large-telescope observing time. 
Second, as long as we do not understand what is causing the photometric variations in magnetic white dwarfs, we cannot fully exclude that the amplitude of the variations depends on the spin period which would introduce a bias in the distribution of periods. 

\begin{table}
    \centering
    \caption{Two-sided Kolmogorov-Smirnov test results for the different samples: all magnetic white dwarfs  (Tables\,\ref{tab:brink} \& and {\it TESS} targets from Table\,\ref{tab:OBS}), {\it TESS} targets from Table\,\ref{tab:OBS}, and white dwarfs within 20\,pc compared to non-magnetic white dwarfs (Table\,\ref{tab:nomag}). The unbiased sample of periods measured with {\it TESS} confirms previous indications that magnetic white dwarfs rotate on average faster than non-magnetic ones. 
    }
    \begin{tabular}{llll}
    \hline 
        Sample 1 & Sample 2 & KS statistic & p-value \\
    \hline
        magnetic & non-magnetic &  0.590 &  3.761$\times 10^{-7}$  \\

        20\,pc & magnetic & 0.333 &  0.300 \\
        {\it TESS} & magnetic &  0.354 &  0.306\\
        20\,pc & non-magnetic & 0.5 &  0.042\\ 
        {\it TESS} & non-magnetic &  0.541 &  0.027\\
    \hline
    Sample & Median period & Average period & St. dev.\\
     & [hr] & [hr] & [hr] \\
    \hline
    non-magnetic & 29.70 & 32.13 & 24.04 \\
    magnetic & 2.44 & 20.23 & 63.17 \\
    20\,pc & 6.19 & 11.61 & 12.22 \\
    {\it TESS} & 6.22 & 10.70 & 10.37 \\
    magnetic (no mergers) & 3.14 & 23.11 & 67.04\\
    \hline
    \end{tabular}
    \label{tab:KS}
\end{table}

\subsection{Relation between rotation and white dwarf mass} 

\begin{figure*}
\includegraphics[width=2.0\columnwidth]{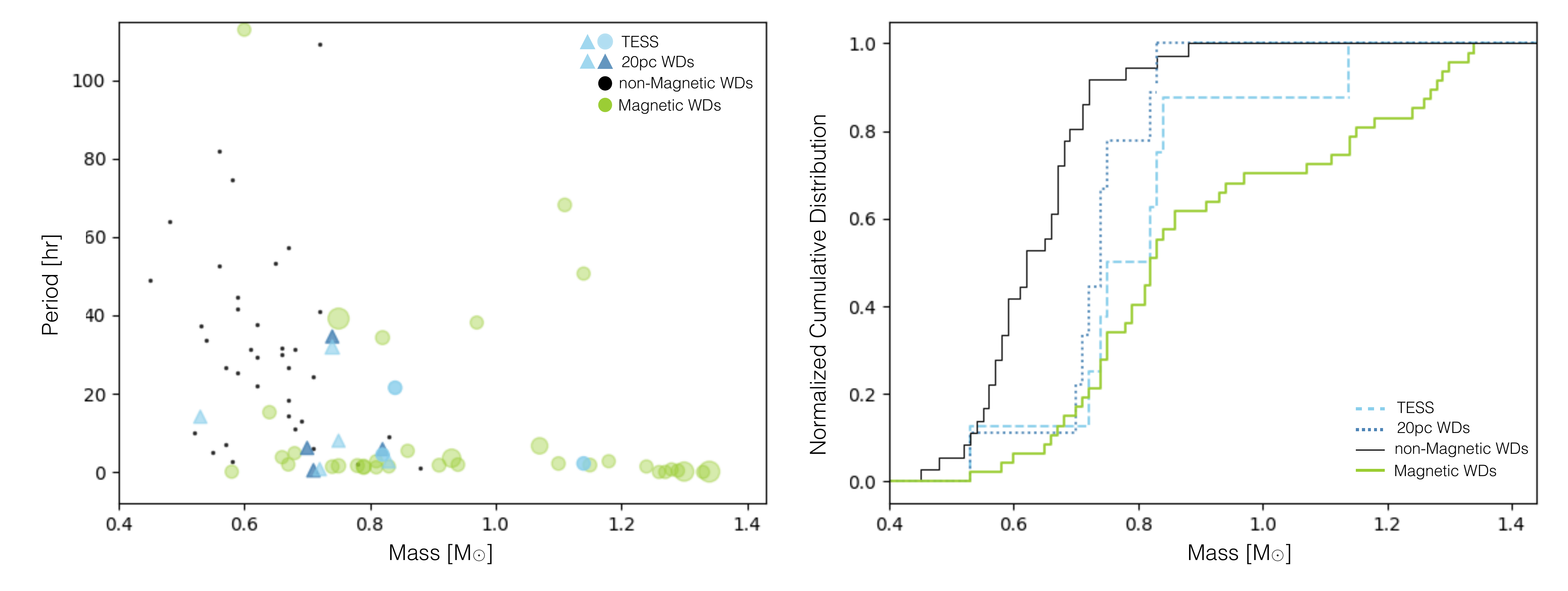}

    \caption{Left: Rotation periods as a function of white dwarf mass for non-magnetic white dwarfs (black dots) from  Table\,\ref{tab:nomag}, 
    and magnetic white dwarfs ( green circles) from Tables\,\ref{tab:brink} and \ref{tab:Localize}. 
    The magnetic white dwarfs with periods measured from {\it TESS} light curves (light blue circles and triangles) and those within 20\,pc (dark blue triangles) have, on average, shorter periods than non-magnetic white dwarfs measured from pulsations. Both samples share the light blue triangles, as some of the white dwarfs in the 20\,pc sample were observed with {\it TESS}. 
    Right: Cumulative distributions of the masses of magnetic and non-magnetic white dwarfs. Both, the {\it TESS}  (dashed light-blue line) and the 20\,pc sample  (dotted dark-blue line) contains slightly more massive white dwarfs than the sample of non-magnetic white dwarfs (solid black line). This difference is statistically significant in both cases ($p=0.0005$ and $p=0.0008$), which is consistent with the fact that magnetic white dwarfs in the 20\,pc sample are slightly more massive than the non-magnetic ones \citep{bagnulo+landstreet21-1}}.

    \label{fig:mass}
\end{figure*}

As a first step to evaluate possible dependencies on the white dwarf mass, we plot the 
cumulative distributions of the white dwarf masses of all previously discussed samples in the right panel of Figure\,\ref{fig:mass}. 
Previously measured spin periods cover a mass range that is very different to that of typical white dwarfs. The mass distributions of this magnetic white dwarf sample  (median: 0.82\,\Msun) and that of the non-magnetic white dwarfs (median: 0.62\,\Msun) are clearly different. 
In contrast, the 20\,pc and the {\it TESS} sample represent the first ones covering spin periods of more typical white dwarf masses  (medians: 0.74\,\Msun\,  and 0.78\,\Msun, respectively).






However, even both these samples show a tendency towards larger white dwarf masses compared to the sample of single non-magnetic white dwarfs. 
This reflects the fact that magnetic white dwarfs in the 20\,pc sample seem to be slightly more massive than non-magnetic white dwarfs as already mentioned by \citet{bagnulo+landstreet21-1}. 
According to KS-tests, this difference in masses 
for both the 20\,pc sample ($p=0.0008$) and the smaller {\it TESS} sample ($p=0.0005$).

In the left panel of Figure\,\ref{fig:mass} we show the rotation period as a function of white dwarf mass for non-magnetic and magnetic white dwarfs. 
In the 20\,pc and {\it TESS} sample we find white dwarfs with masses exceeding $0.75$\,\Msun to have short periods very similar to what \citet{kawaler15-1} 
and \citet{hermesetal17-1} found for the only three non-magnetic white dwarfs in this mass range. 

However, as we shall see, a tendency towards shorter spin periods for larger masses seen in the sample of non-magnetic white dwarfs (but indicated by just three systems) is in general not obvious in magnetic white dwarfs. 
For white dwarfs exceeding $0.85$\,\Msun, spin periods are only known for magnetic white dwarfs. 
Several very massive and magnetic white dwarfs rotate fast, i.e. with spin periods of minutes. 
Such typically larger masses and short rotation periods are expected for white dwarf mergers 
\citep{schwab21-1} and are entirely absent in the 20\,pc sample, i.e. they are rare and largely over-represented in Figure\,\ref{fig:mass}. 
There are, however, also four magnetic white dwarfs with masses above 
$0.85$\,\Msun and spin periods of the order of days (one of them is not shown in the figure because its period is longer than 400\,hours). 

\begin{figure*}
    \includegraphics[width=2.0\columnwidth]{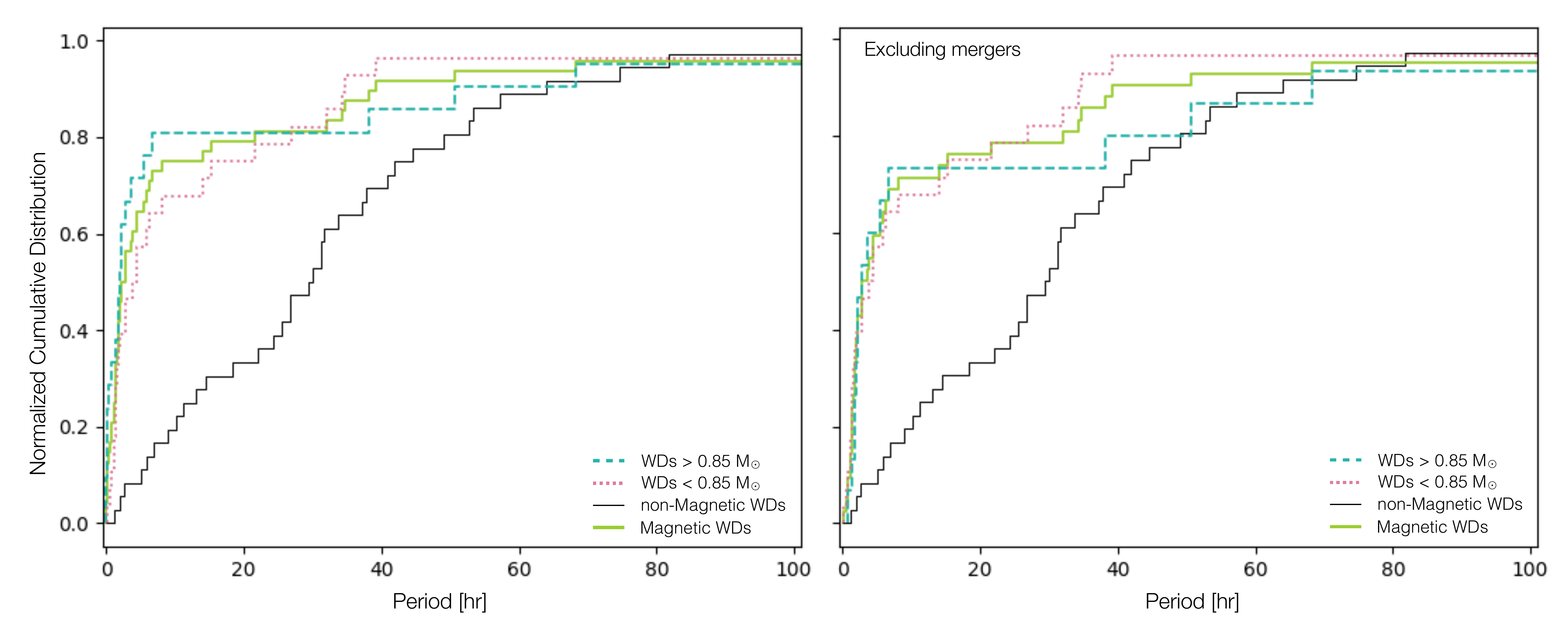}
    \caption{
     Cumulative period distributions of magnetic and non-magnetic white dwarfs. {\it Left:} non-magnetic white dwarfs (solid black line) are compared with the magnetic ones separated in three different samples: all magnetic white dwarfs (Tables\,\ref{tab:brink} and \ref{tab:OBS}, solid green line), magnetic white dwarfs with masses below 0.85\,\Msun (dotted pink line), and massive magnetic white dwarfs ($\geq$\,0.85\,\Msun, dashed blue line). {\it Right:} The same samples as in the left panel, but removing white dwarfs that very likely formed throughout mergers (rotational periods shorter than 0.38\,hr and masses larger than 1.25\,\Msun). No indications for a relation between spin period and white dwarf mass can be identified especially if white dwarfs that likely formed through mergers are eliminated.
    }
    \label{fig:CDF1}
\end{figure*}

In order to test for dependencies of the spin periods on white dwarf masses, we again used cumulative distributions and the KS-test. In the left panel of Figure\,\ref{fig:CDF1} we separate the magnetic white dwarfs into samples of normal mass white dwarfs ($\leq0.85$\,\Msun) and massive white dwarfs ($>0.85$\,\Msun) and compare the cumulative distributions. 
In the right panel of Figure\,\ref{fig:CDF1} we compare the same distributions but this time we excluded the obvious merger products
which reduces the number of stars in the massive white dwarf sample and the full sample of magnetic white dwarfs by six. There is no significant difference in spin periods for different masses 
(p-values for all comparisons exceed $p=0.33$). This result does not depend on the (admittedly arbitrary) value at which we separate high and low-mass white dwarfs. Using e.g. $0.75$\,\Msun or $0.8$\,\Msun instead of $0.85$\,\Msun provides essentially the same results.
However, we again advocate caution with the above results because we are dealing with low-number statistics. More spin period measurements are clearly needed before solid conclusions can be drawn.

\subsection{Dependencies on spectral type} 

DZ white dwarfs show metal absorption lines which are indicative of ongoing (or recent) accretion events \citep[e.g..][]{jura03-1,gaensicke06, Vanderburg15, gaensicke19}.  
Intriguingly, \citet{Kawka19} and \citet{hollandsetal17-1} discovered a large number of magnetic white dwarfs among samples of cool metal-polluted white dwarfs and claimed that there might be an increase in magnetism among these white dwarfs
\citep[see also][]{bagnulo+landstreet19-1}. This alleged increase in magnetism among metal-polluted white dwarfs motivated \citet{schreiberetal21-2} to speculate that faster rotation, generated by the accretion of planetary debris, could trigger the crystallization and rotation-driven dynamo. 
Indeed, accretion can easily reduce the spin period from a few days to a few hours \citep[][their Figure 2]{schreiberetal21-2}.

However, there are two fundamental arguments against a relation between metal pollution and magnetism. 
First, and most importantly, the volume-limited sample provided by \citet{bagnulo+landstreet21-1} shows that there most likely is no increase in magnetism for metal-polluted white dwarfs. Of the 26 metal-polluted white dwarfs within $20$\,pc of the Sun, only six are magnetic. This corresponds to a fraction of $0.23\pm0.08$ magnetic white dwarfs among metal-polluted white dwarfs which is identical to the overall fraction of magnetic white dwarfs among cool ($\Teff \leq \,15\,000$\,K) white dwarfs in the 20\,pc sample ($33/139=0.24\pm0.035$). In other words, there are no indications for an increase in magnetism among metal-polluted white dwarfs in this volume limited sample. 

Second, if a cold and old DA white dwarf is observed today, accretion of planetary debris onto this white dwarf might have occurred in the past and therefore could have increased the white dwarf's rotation rate. 
It is therefore unclear if metal-polluted white dwarfs should indeed rotate faster on average than white dwarfs of other spectral types. If the accretion of planetary debris occurs in relatively short episodes on a large number of white dwarfs over a long period of time, one would rather expect a relation between rotation rate and age. 

We here present measurements of the rotation periods of metal-polluted white dwarfs which complement the rough estimate of $18.5$\,hr for WD\,G29-38 by \citet{Thompson10}. 
The two clearly variable light curves we measured with SPECULOOS 
show that these two metal-polluted white dwarfs have periods of the order of a few hours. 
Our SPECULOOS observations show that LSPM\,J0107+2650 has a period of $4.8$\,hr while for WD\,2138-332 we estimate a period in the range of $4-12$\,hr. 
As we observed several magnetic DZ white dwarfs with SPECULOOS the two measurements we performed with SPECULOOS are potentially biased towards short periods. Given that in addition both targets have periods typical for magnetic white dwarfs, we conclude that the currently available data does not provide evidence for faster rotation among metal-polluted white dwarfs (nor against it). 

\begin{table*}
    \centering
    \caption{ Stellar parameters from {\it TESS} and SPECULOOS targets obtained from literature. The white dwarf spectral types, temperatures, masses,  log\,g,   cooling age and magnetic field magnitude have been taken from \citet{bagnulo+landstreet21-1}  and \citet{bagnulo+landstreet22-1} for nine of our targets.  The masses for the {\it TESS} white dwarf with distance  larger tan 20\,pc were obtained from \citet{Gentile19}. For the metal-polluted white dwarf LSPM\,J0107+2650 whose parameters are listed in \citet{hollandsetal17-1}, and for WD\,2359-434 whose parameters where obtained form \citet{bagnulo+landstreet19-1}. The listed distances were obtained from the {\it Gaia} catalog \citep{gaiaDistance21}.  The targets are sorted based on the measured amplitude of their light curves.}
	\label{tab:OBS}
    \begin{tabular}{llllllllll}
    \hline
    Name & Type & Temperature & Mass & log g & Age & Magnetic field & Distance &  Period  & Normalize \\
      &  & [K] & [M$_{\odot}$] &  & [Gy]  & [MG] & [pc]  & [hr]  & amplitude  \\
    \hline
    {\it TESS} &&&&&&\\
    \hline
    WD\,0011-134 &   DAH & 5855 & 0.72  & 8.22 & 4.1 &12   &  $18.56 \pm 0.01$ &    $0.736 \pm 0.007$  & 0.0033$\pm$0.0004 \\
    WD\,0009+501 &   DAH & 6445 & 0.75 & 8.25 & 3.24 & 0.25 & 10.87 $\pm$ 0.01  &  4.0086 $\pm$ 0.0001 & 0.0034 $\pm$ 0.0003  \\
    WD\,2359-434 &   DAH & 8390 & 0.83 & 8.37 & 1.83 & 0.10 & $8.33 \pm 0.01$   &    2.694 $\pm 0.002$  & 0.0037 $\pm$ 0.0001\\
    WD\,0011-721 &   DAH & 6340 & 0.53 &7.89 & 1.66  & 0.37 & $18.79 \pm 0.01$  &   $14.13 \pm 0.38$ & 0.0038 $\pm$ 0.0003\\
    WD\,J075328.47–511436.98 &   DAH & 9280 &  0.84 & 8.39 & - & 19 &  32.71$\pm$0.02  &  21.51$\pm$ 0.01 & 0.0062 $\pm$ 0.0005  \\
    WD\,J171652.09–590636.29 &   DAH & 8600 &  0.82 & 8.37 & - & 0.7 & 29.83$\pm$0.03  &  4.4398$\pm$0.0003 & 0.0080 $\pm$ 0.0001\\
    WD\,0912+536 &   DCH & 7170 & 0.74 & 8.27 & 2.48& 100  & $10.27 \pm 0.01$  &   $31.93 \pm 0.13$ & 0.0189 $\pm$ 0.0001 \\
    WD\,0041-102 &  DBAH & 21341 & 1.14 & - &0.35 & 20 & 31.1$\pm$0.041 &  2.19$\pm$ 1.09   & 0.0327 $\pm$ 0.0003    \\
    \hline
    SPECULOOS &&&&&&\\
    \hline
    WD\,2138-332 &   DZH & 6908 & 0.6 &8.05 & 1.72 & 0.05 & $16.11 \pm 0.01$ &  6.19  $\pm$ 0.05 & 0.008 $\pm$ 0.001\\
    LSPM\,J0107+2650 &  DZ & 6190 & 0.68 & -&2.9 & 3.37 & $71.96 \pm 1.16$ &   4.82 $\pm$ 0.29 & 0.113 $\pm$ 0.01 \\  
    \hline
    \end{tabular}
 For the white dwarf WD\,0009+501, the table shows the spin period measured by {\it TESS}, which is half of the real spin period previously obtained by spectropolarimetric observations in \citet{valyavinetal05-1}.
\end{table*}

While there seems to be no strong evidence for a dependence of the spin period on the spectral types 
we observe indications for a relation between 
spectral type and amplitude of the detected photometric variations. 
In Table\,\ref{tab:Localize} we list the amplitudes of the measured variations for the white dwarfs presented in this work.  In Table\,\ref{tab:OBS}, we list the white dwarfs according to their amplitudes, from smallest to largest, along with their stellar parameters from the literature.

While the four DA white dwarfs show variations with an amplitude of $\sim0.3-0.8$ per cent, the DZ and DC white dwarfs in our sample show variations with larger amplitudes ranging from $0.8$ to $10$ per cent. 
If confirmed by analyzing larger samples, this finding might indicate that the variations are produced by different effects in the different spectral types.

\section{Discussion}

We presented periodic light curve variations of  eleven magnetic white dwarfs and interpret the periodicity as reflecting the spin period of the white dwarfs. We compared these measurements with spin periods of magnetic and non-magnetic white dwarfs from the literature and obtain the following results: 
%
(i)  The spin periods of magnetic white dwarfs are shorter than that of non-magnetic white dwarfs but larger unbiased samples are needed to confirm this. 
We note that the spin periods have been measured in an unbiased way for only  six of the 33 magnetic white dwarfs within $20$\,\,pc, and therefore definite conclusions cannot be drawn. 
%
(ii) Massive ($\gappr \,0.85\,\mathrm{M}_{\odot}$) magnetic white dwarfs rotate on average slightly faster  than lower-mass magnetic white dwarfs. However, this difference disappears if strong merger candidates (very fast-rotating massive white dwarfs) are eliminated from the sample. 
A general trend for decreasing rotation periods with white dwarf mass can therefore not be established. Only the obvious merger candidates, i.e. very massive ($>1.25$\Msun) and fast rotating (of the orders of minutes) white dwarfs stand out. 
%
(iii) We estimated rotation rates for two metal-polluted white dwarfs and find that both of them have periods of a few hours. Given that the previously published estimate of $18.5$\,h for
WD\,G29-38 \citep{Thompson10} can be considered a rather rough estimate, these are the first robust measurements of rotation periods of metal-polluted white dwarfs. 
(iv) We found the amplitudes of the variations to depend on the spectral type of the white dwarfs, with variations in DA white dwarfs showing smaller amplitudes than those of DZ or DC white dwarfs. 
However, a larger homogeneous sample of light curves of magnetic white dwarfs is needed to confirm this potential trend.

In what follows we discuss how robust our findings are and relate them to theories suggested for the origin of magnetic fields in white dwarfs.

\subsection{Spin periods and the crystallization dynamo} 

Investigating the origin of magnetic fields in white dwarfs in close binary stars and based on the early work by \citet{isernetal17-1}, \citet{schreiberetal21-1} assumed fast rotation to be a necessary ingredient for the dynamo to work. The proposed scenario assumes that white dwarfs in detached post-common envelope binaries are born without a magnetic field and that only when a crystallizing white dwarf is spun-up by accretion of angular momentum in semi-detached cataclysmic variables does the crystallization and rotation-driven dynamo generates a strong magnetic field. If the field is strong enough, it can connect with the field of the secondary star and transfer spin angular momentum to the orbit. This can cause the binary to detach for a relatively short period of time. 
The proposed scenario is very attractive as it can explain several otherwise inexplicable observations: the absence of strongly magnetic white dwarfs in young detached post common envelope binaries \citep{lieberetal03-1}, the population of old and close to Roche-lobe filling post common envelope binaries \citep{parsonsetal21-1}, the existence of the fast spinning white dwarf radio pulsar AR\,Sco \citep{marshetal16-1}, the absence of X-ray bright polars in globular clusters \citep{bellonietal21-1}, and the fact all but one white dwarf in close detached double white dwarfs binaries are not magnetic 
\citep{schreiberetal22-1}. 

However, as we have mentioned previously, the only magnetic white dwarfs that rotate with spin periods of the order of minutes are very massive and are consistent with being formed through stellar mergers. This implies that the fast rotation suggested by \citet{isernetal17-1} and \citet{schreiberetal21-1} is very unlikely to generally play a role in the magnetic field generation. At the very best fast rotation may increase the strength of generated fields.

Recently, \citet{ginzburgetal22-1} showed that convection in crystallizing white dwarfs is slower than previously estimated by \citet{isernetal17-1} which translates to slower rotation rates being required 
for the dynamo to work. \citet{ginzburgetal22-1} also estimated the time scale for the generated magnetic field to appear on the surface of the white dwarf to be between $\simeq0.1-1$\,Gyr depending on the extension of the carbon enriched convection zone. 
The results presented in this paper might be consistent with the slow-convection dynamo playing a role in the magnetic field generation in white dwarfs.  We found that seven of nine periods measured with {\it TESS} of magnetic white dwarfs rotate with periods of just a few hours and/or have only relatively weak magnetic fields which could be consistent with the slow convection dynamo. 

However, it also becomes immediately clear that the dynamo cannot be responsible for all magnetic fields in white dwarfs. In particular, 
the DCH white dwarf WD\,0912+536 has a very strong fields ($\gappr\, 100$\,MG), a typical white dwarf mass, and a spin periods significantly longer than one day (similar to that of longer period non-magnetic white dwarfs). Furthermore, the DAH white dwarf  WD J075328.47–511436.98 rotates with a spin period exceeding 20 hours and hosts a magnetic field with a strength of 19\,MG which is also difficult to explain with the dynamo \citep[see][their Figure 4]{ginzburgetal22-1}.  

\begin{figure}
\includegraphics[width=1.0\columnwidth]{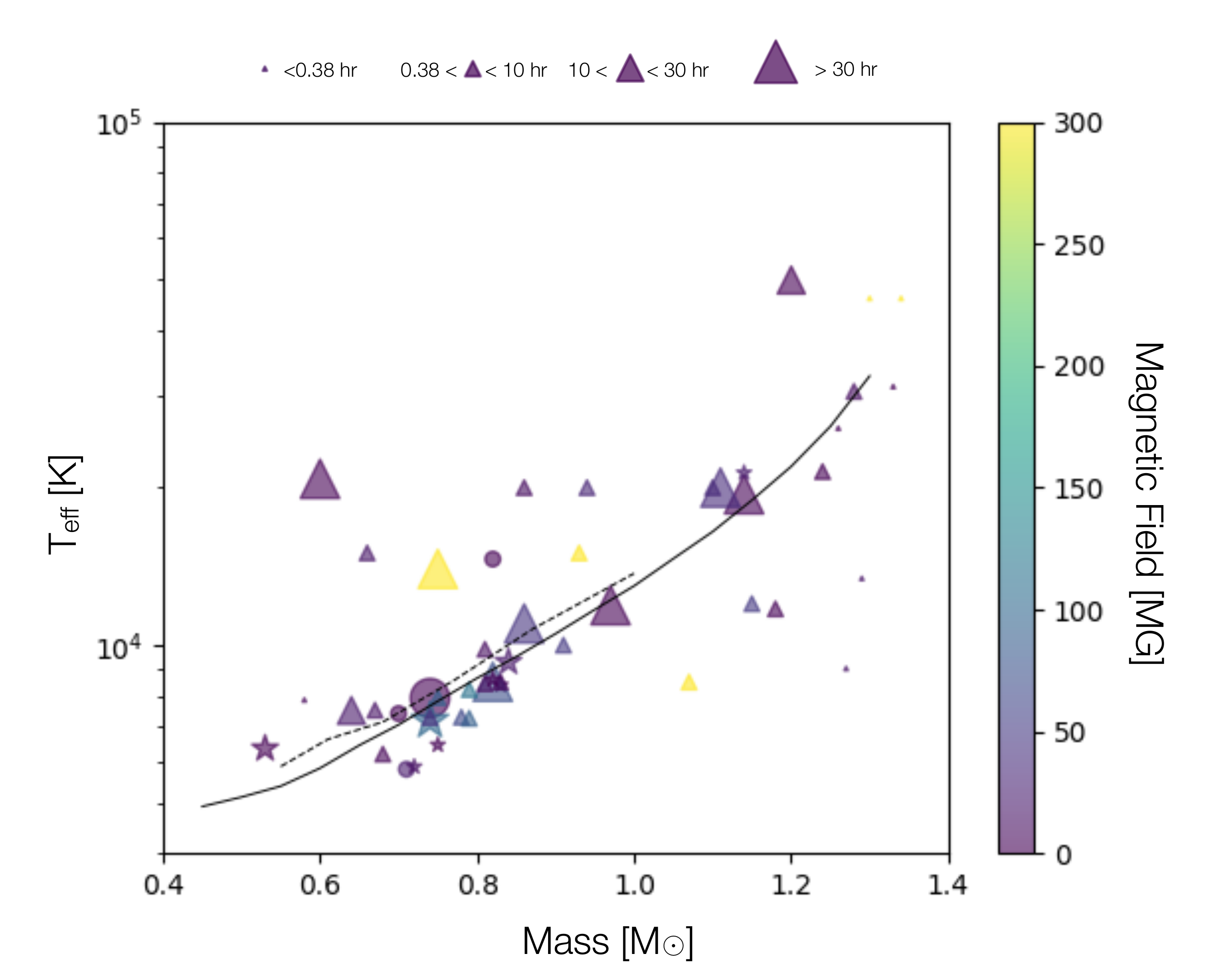}

    \caption{Mass-temperature distribution of magnetic white dwarfs. White dwarfs with periods measured from {\it TESS} light curves are represented by stars, white dwarfs with measured periods and within 20\,pc  from Table\,\ref{tab:brink} are marked as filled circles, and the white dwarfs from Table\,\ref{tab:brink} are represented by triangles. Colours show the magnetic field strength with a maximum of 300\,MG while the size of the objects (scale on the top of the image) represents different ranges of the rotational period of the white dwarfs. The solid and dashed lines provide the temperature limits for the onset of crystallization according to \citet{Bedard20} and \citet{Salaris10}, respectively.}
    \label{fig:mass_temp}
\end{figure}

To further investigate to which degree the proposed dynamo may contribute to the magnetic field generation in white dwarfs, 
we compare the available information, white dwarf mass, rotation period, field strengths, and effective temperature of magnetic white dwarfs with the predictions of the dynamo scenario in fig.\,\ref{fig:mass_temp}.  
If the crystallization and rotation-driven dynamo represents an important channel for magnetic field generation in white dwarfs, one would expect to see an accumulation of magnetic white dwarfs with crystallizing cores rotating relatively fast or a relation between rotation period and field strength. 

Indeed, most magnetic white dwarfs have periods of a few hours or less, magnetic field strength of less than a few MG and accumulate in the lower left corner  
of Fig.\,\ref{fig:mass_temp}, i.e. have low temperatures and typical white dwarf masses. It also seems that a significant number of magnetic white dwarfs cluster around the onset of crystallization. Although some of them rotate too slowly to generate their magnetic field through the dynamo, this clustering might imply that crystallization plays a role in the magnetic field generation of some white dwarfs.   
Also the magnetic white dwarfs in the {\it TESS} and 20\,pc sample (stars and circles) are mostly close to the crystallization limit but some of them rotate very slowly yet host strong fields.   
If we include magnetic white dwarfs from the literature (triangles), it further becomes obvious that magnetic white dwarfs rotating relatively fast can be found above the temperature limit for the onset of crystallization as well as below. 
This suggests that the dynamo might be responsible for a fraction of the observed magnetic fields in single white dwarfs but is clearly unable to explain all of them.

\subsection{Fast spinning massive white dwarfs and the merger channel}

The only crystal clear tendency that can be seen in fig.\,\ref{fig:mass_temp} is the existence of very fast rotators among high-mass white dwarfs, i.e. we see evidence for the merger channel \citep{garcia-berroretal12-1} producing magnetic white dwarfs. 

Our literature search provided a relatively large number of high-mass white dwarfs with short spin periods. 
These white dwarfs are absent in the 20\,pc sample \citep{bagnulo+landstreet21-1} and are therefore rare \citep{Gentile21}.
There is an increasing amount of evidence that these magnetic high-mass white dwarfs are the product of mergers as suggested early by \citet{garcia-berroretal12-1}. 
\citet{kilicetal23-1} showed that $40$ per cent of high-mass white dwarfs are magnetic which exceeds the 20 per cent found in volume-limited samples \citep{bagnulo+landstreet21-1}. 
This finding is in agreement with \citet{bagnulo+landstreet22-1} who find that in high-mass white dwarfs, magnetic fields are extremely common, very strong and appear immediately in the cooling phase while lower-mass white dwarfs become magnetic close to the onset of crystallization. Further evidence for the generation of magnetic fields through mergers is provided by \citet{pelisolietal22-1} who found evidence for the three known magnetic hot sub-dwarfs to be the product of a merger.

Theoretical simulations of the merger process predict the remnant to have very short spin periods of just $\simeq10-20$ minutes \citep{schwab21-1}. 
This prediction fits with the most massive white dwarfs that are rotating with periods of this order (we used a rather arbitrary limit of $0.38$\,hr in fig.\,\ref{fig:mass_temp}). We, therefore, conclude that at least these fast-spinning, strongly magnetic, and very massive white dwarfs are the product of mergers.  
However, it seems that not all mergers produce strongly magnetic white dwarfs. \citet{Hollands20} found a massive white dwarf that is likely the product of a merger and derived an upper limit on the field strength of only 50\,kG.

The magnetic white dwarfs in the mass range of $\sim0.8-1.2$ that do not spin with periods of minutes, show spin periods similar to that of less massive magnetic white dwarfs. This means, the spin periods of these white dwarfs alone do not provide evidence for them emerging from a different channel than the lower mass white dwarfs. In other words, the spin periods alone do not provide indications that these white dwarfs might be mergers as well.

\subsection{Constraints on magnetic field generation in white dwarfs} 

Of the mechanisms suggested for magnetic field generation, the merger channel is the most convincing and the only one that can count on clear observational evidence. However, only very few white dwarfs are produced through this channel which is clearly illustrated by the complete absence of such white dwarfs in the 20\,pc sample \citep{bagnulo+landstreet21-1}. 

Concerning the dominant population of magnetic white dwarfs we know the following: the 20\,pc sample 
established by \citet{bagnulo+landstreet21-1} shows that in general magnetic fields in white dwarfs appear late, i.e when the white dwarf 
has cooled to temperatures similar to the onset of crystallization. This is illustrated by the white dwarfs marked as stars and circles in Figure\,\ref{fig:mass_temp}. In this work, we complemented this result by finding that a significant fraction of these cool nearby white dwarfs rotate fast (periods of a few hours) while others are slowly rotating (longer than 20\,hours). 
In addition, we find that on average magnetic white dwarfs rotate faster than non-magnetic white dwarfs. These constraints seem to imply that in general rotation plays a role in the generation of white dwarf magnetic fields but not in all cases. 

 It could be not just one additional channel to the merger channel but several, and only one of them necessitates relatively fast rotation.
The generally late appearance of the magnetic fields could imply that crystallization plays a role or that the fields appear late for a different reason, e.g. because they were buried following the white dwarf formation. 
We clearly need a larger sample of well characterized magnetic white dwarfs and new theoretical ideas to finally understand the formation of magnetic fields in white dwarfs.  

\subsection{Magnetic white dwarfs in binaries}

The absence of strongly magnetic white dwarfs in young post-common envelope binaries was one of the main motivations for the development of the crystallization and rotation-driven dynamo \citep{schreiberetal21-1}. This absence of young magnetic white dwarfs in close binaries is consistent with the absence of strongly magnetic single white dwarfs in  the 20\,pc sample \citep{bagnulo+landstreet21-1}. Whether the generally late appearance of magnetic fields in white dwarfs is caused by a dynamo that requires crystallization or if the magnetic field appears late for another reason remains an open question.

In any case, whether the magnetic field is generated late due to a dynamo (that does depend on rotation) or appears late in the evolution of a white dwarf for another reason (e.g. because it was buried), the evolutionary sequence suggested by \citet{schreiberetal21-1} for white dwarf binaries (i.e. magnetic field appearance/generation during the CV phase, connection with the field of the secondary, angular momentum transfer to the orbit, cessation of mass transfer, synchronization, re-start of mass transfer) remains a promising scenario to explain the observations of magnetic white dwarfs in post common envelope binaries and CVs.  

The situation for double white dwarfs is similar. As shown by \citet{schreiberetal22-1}, only one magnetic white dwarf is known in these systems (which is close to the crystallization boundary and rapidly rotating). If rotation does not play a crucial role in the magnetic field generation, cool white dwarfs in close double white dwarfs should develop magnetic fields. In the 20\,pc sample the occurrence rate of magnetism increases for effective temperatures in the range of $4\,000-8\,000$\,K \citep[][thier Figure 2]{bagnulo+landstreet21-1} while the vast majority of both components in double white dwarfs are hotter. It might therefore be that in both populations, close double white dwarfs, and post common envelope binaries with main sequence star companion, the late appearance of magnetic fields has not yet been observed. 

Further searches for magnetism in cool white dwarfs that are part of post-common envelope binaries or double white dwarfs would significantly help to settle open questions. If rotation is not a key ingredient for the appearance of magnetic fields, we should find magnetic white dwarfs in detached double white dwarfs and among post-common envelope binaries that contain cool white dwarfs and that are not close to Roche-lobe filling. 
The latter does not seem to be the case \citep{parsonsetal21-1}, but the sample size is still small and heavily biased towards systems that are close to Roche-lobe filling as the magnetic fields are detected through cyclotron emission via wind accretion. 
Further observations, especially of systems containing the coolest white dwarfs are certainly required (but might be challenging).

\section{Conclusions}

The 20\,pc sample provided by \citet{bagnulo+landstreet21-1} showed that white dwarfs become magnetic when they are older, typically with ages exceeding 2\,Gyr \citep[see also][]{bagnulo+landstreet22-1}, reaching effective temperatures close to the crystallization limit. While detached post-common envelope binaries are hardly ever magnetic, a large number of their descendants, semi-detached cataclysmic variables turned out to host a magnetic white dwarf. 
In other words, there is clear observational evidence for the late appearance of magnetic fields in white dwarfs. 
Based on this finding and the early work by \citet{isernetal17-1}, a crystallization and rotation-driven dynamo has been proposed for the origin of magnetic fields in white dwarfs \citep{schreiberetal21-1,ginzburgetal22-1}.   

 We presented measurements of the spin periods of  eight magnetic white dwarfs using {\it TESS} light curves and found that their spin periods are similar to those of previously identified magnetic white dwarfs and shorter than that of non-magnetic white dwarfs. 
Four of the periods we measured agree with previously measured periods and in one of them the {\it TESS} period is half the real period.  
One white dwarf in our sample has magnetic field strengths of $\sim100$\,MG but a long rotation rate of $32$\,hr. 
This result indicates that fast rotation plays a role in the magnetic field generation in general but is not always required. 

In a sample of magnetic white dwarfs compiled from the literature, some massive white dwarfs entirely consistent with being merger products stand out. In the remaining systems, we do not find evidence for a dependence of spin period on white dwarfs mass but a larger and/or complete volume-limited sample of magnetic white dwarfs is needed to confirm this impression.  

The crystallization and rotation-driven dynamo might be responsible for a fraction of the magnetic fields in white dwarfs. If it is, a third mechanism to the dynamo and mergers would probably be required. Alternatively, the proposed dynamo does not play a role but just one channel (complementing the magnetic white dwarfs resulting from mergers), that for some reason leads to the late appearance of magnetic fields and on average shorter spin periods, is responsible for the generation of most magnetic fields in white dwarfs.  
%
Independent of the reasons for the late appearance of magentic fields, 
the new evolutionary sequence suggested by \citet{schreiberetal21-1} for white dwarfs in binaries remains plausible.


\section*{Acknowledgements}

 We thank the anonymous referee for their comments, which have helped us to improve the paper. We thank the Kavli Institute for Theoretical Physics (KITP) for hosting the program "White Dwarfs as Probes of the Evolution of Planets, Stars, the Milky Way and the Expanding Universe". 
This research was supported in part by the National Science Foundation under Grant No. NSF PHY-1748958.
MSH acknowledges support through a postdoc fellowship from the DGIIE (General Direction of Research, Innovation and Undertaking) of the UTFSM. MRS acknowledges support by ANID, – Millennium Science Initiative Program – NCN19\_171, and  FONDECYT (grant 1221059). OT was supported by FONDECYT grant 3210382. JDL acknowledges the financial support of the Natural Sciences and Engineering Research Council of Canada (NSERC), funding reference number 6377-2016. SGP acknowledges the support of a STFC Ernest Rutherford Fellowship. 
For the purpose of open access, the author has applied a creative commons attribution (CC BY) licence to any author accepted manuscript version arising. 
\section*{Data Availability}

SPECULOOS photometry is available at the data archive of the European Southern Observatory under ESO programme 0108.A9010(C) and the analyzed {\it TESS} data is publically available.



\bibliographystyle{mnras}
\bibliography{TESS_magWD} 





\appendix

\section{Details on the 20 pc {\it TESS} sample}

\subsection{{\it TESS} sectors}

We provide the lists of {\it TESS} sectors that we analyzed for all targets within 20\,pc. This includes white dwarfs with clear photometric variations  (see Table\,\ref{tab:secotrs20}).  

\begin{table*}
    \centering
        \caption{List of  magnetic white dwarfs with {\it TESS} observations from the 20\,pc sample. Targets for which we found clear periodic variations are marked with a \checkmark symbol. 
        For these systems, we provide the identification number for all the analyzed {\it TESS} sectors. For those targets without clear variations (marked with $\times$) we only provide the total number of {\it TESS} sectors available for each white dwarf. Values in the parenthesis indicate the contamination level for the corresponding sector .}
    \label{tab:secotrs20}
    \begin{tabular}{llllll}
    \hline 
      Name& TIC name &Variation & Distance [pc]  & FluxCT & Exposure time [s]   \\
    \hline
   WD\,0009+501& TIC\,201892746 & \checkmark &   10.87 &  17(0.0\,\%), 57(4.7\,\%) & 120, 20  \\ 
   WD\,0011-134& TIC\,289712694& \checkmark & 18.55 &  29(0.0\,\%) & 120 \\
   WD\,0011-721 & TIC\,328029653 & \checkmark  & 18.79 &  1(10.8\,\%), 27(8.2\,\%), 28(10.4\,\%) & 120, 120, 120  \\ 
   WD\,0912+536 &  TIC\,251080865 & \checkmark  & 10.27 &  21(0.0\,\%), 47(0.0\,\%)  & 120, 20   \\
   WD\,2359-434 & TIC\,321979116 &\checkmark  & 8.33 &  2(0.1\,\%), 29(0.1\,\%) & 120, 20 \\ 
   
    \hline
    WD\,0004+122&  TIC\,357353518 &$\times$ & 17.45  &    2(0.0\,\%) \\ 
    WD\,0503-174&  TIC\,169379648  &$\times$ & 19.34 &   2(11.1\% ) \\
    WD\,0121-429&  TIC\,262548040 &$\times$ & 18.45  &  4(8.4\,\%) \\
    WD\,0233-242&  TIC\,65324009 &$\times$  & 18.46  &  2(3.4\,\%)\\
    WD\,0322-019& TIC\,279198715  &$\times$ & 16.93  &    2(3.1\,\%) \\
    WD\,0548-001&  TIC\,176670072 &$\times$ & 11.22  &   2(1.7\,\%) \\
    WD\,0708-670&  TIC\,300013123 &$\times$ & 16.96  &   22(48.9\,\%) \\
   WD\,0810−353 &  TIC\,145863747 &$\times$ & 11.17 &  1\,(28.3\,\%) \\
   WD\,0816−310 &  TIC\,147018085  &$\times$  & 19.36 &  1\,(7.6\,\%) \\
    WD\,1008+290 & TIC\,241190677 & $\times$ & 14.73  &   2(0.0\,\%) \\
   WD\,1009-184 & TIC\,52104043& $\times$  & 18.08  &    2(1.3\,\%)   \\
    WD\,1036-204 & TIC\,179299789 &$\times$  & 14.11 &  2(35.0\,\%) \\
    WD\,1309+853 &  TIC\,154903874  &$\times$ & 16.47&   10(10.9\,\%)  \\ 
    WD\,1315-781 &  TIC\,448023860  &$\times$ & 19.28 &   4(28.8\,\%) \\
    WD\,1532+129 &  TIC\,157110489  &$\times$ & 19.25 &   1(8.4\,\%) \\ 
    WD\,1748+708 &  TIC\,233212451  &$\times$ &  6.21 &   25(0.0\,\%) \\ 
   WD\,1829+547 & TIC\,390019679 &$\times$ & 17.03 &   1\,(0.0\,\%) \\
   WD\,1900+705 &  TIC\,229797408 &$\times$ & 12.87 &   1\,(0.0\,\%)\\
    WD\,2047+372 &  TIC\,390019679  &$\times$  & 17.58&  3(22.9\,\%) \\
    WD\,2105-820 &  TIC\,403995834  &$\times$ & 16.17 &   3(5.5\,\%) \\
    WD\,2150+591 & TIC\,283414280   &$\times$ & 8.47 &   2(68.0\,\%)\\
   WD\,2153-512 & TIC\,140045537 &$\times$  & 14.85 &  2(47\,\%)\\
     \hline
    \end{tabular}
\end{table*}

\begin{table*}
    \centering
        \caption{List of  magnetic white dwarfs with {\it TESS} observations from the 40\,pc sample. Targets for which we found clear periodic variations are marked with a \checkmark symbol. For these systems, we provide the identification number for all the analyzed {\it TESS} sectors. For those targets without clear variations (marked with $\times$) we only provide the total number of {\it TESS} sectors available for each white dwarf. Values in the parenthesis indicate the contamination level for the corresponding sector.}
    \label{tab:secotrs40}
    \begin{tabular}{llllll}
    \hline 
      Name&  TIC name  & Variation & Distance [pc] & FluxCT & Exposure time [s]   \\
    \hline
   WD\,0041−102 & TIC\,3888273& \checkmark & 31.1 &  3(0.3\,\%), 30(0.3\,\%) & 120, 20\\
   WD\,0232+525 & TIC\,249952539  &\checkmark & 28.8 & 18(22.3\,\%), 58(5.2\,\%) & 120, 20\\   
   WD\,J075328.47--511436.98 & TIC\,269071459 &\checkmark  &  32.71   &34(59.7\,\%), 35(21.7\,\%), &120, 120,  \\
    & & & & 36(29.7\,\%), 61(17.4\,\%) &  120, 20\\

    \hline

    WD\,0316−849 &  TIC\,267166357 &  $\times$ & 29.38   &4(36.8\,\%) \\
    WD\,0945+245 & TIC\,98413819 &  $\times$ & 36.20     &4(1.4\,\%) \\
    WD\,1658+440 &  TIC\,115613388  &  $\times$ & 31.65  &5(4.3\,\%) \\
    WD\,1704+481.1 &  TIC\,274677205 & $\times$ & 39.45&   5(50.2\,\%) \\ 
    WD\,2010+310 & TIC\,92633917  &  $\times$ & 30.76 &   4(56.6\,\%) \\ 
   WD\,1008−242 & TIC\,168071263  &$\times$ & 39.7    &1(0.0\,\%)\\ 
   WD\,J091808.59--443724.25 & TIC\,75823453 & $\times$  & 28.35  &  2(65.1\,\%)  \\
   WD\,J094240.23--463717.68 & TIC\,33724884  &$\times$  & 20.47    &36(65.5\,\%)  \\
   WD\,J171652.09--590636.29 & TIC\,380174982  &$\times$  & 29.83     &2(74.7\,\%)   \\ 
    WD\,J001830.36−350144.71 & TIC\,63695966 & $\times$ & 35.65    &1(4.9\,\%) \\
    WD\,J014240.09−171410.85 & TIC\,404466241 & $\times$  & 40.0  &1(4.0\,\%) \\
    WD\,J025245.61−752244.56 & TIC\,426018482 & $\times$  &  31.19   &6(20.3\,\%) \\
    WD\,J035531.89−561128.32 & TIC\,197909428 & $\times$  & 32.95    &3(0.0\,\%) \\
    WD\,J042021.33−293426.26 & TIC\,179107237 & $\times$  & 31.09     &2(2.5\,\%) \\
    WD\,J050552.46−172243.48 & TIC\,169379648 & $\times$  & 19.34    &2(11.1\,\%) \\
    WD\,J101947.34−340221.88 & TIC\,71407717 & $\times$  & 27.54    &1(17.1\,\%) \\
    WD\,J103706.75−441236.96 & TIC\,146587982 & $\times$  & 39.10   &1(36.3\,\%) \\
    WD\,J104646.00−414638.85 & TIC\,106989158 & $\times$  & 28.23    &2(13.4\,\%) \\
    WD\,J121456.38−023402.84 & TIC\,349446042 & $\times$  & 38.04    &2(0.0\,\%) \\
    *WD\,J140115.27−-391432.21&  TIC\,179029240  &$\times$  & 27.78   & 1(38.6\,\%) \\
    WD\,J180345.86−752318.35  &  TIC\,292631204  &$\times$  & 31.29  & 1(54.9\,\%) \\
    WD\,J193538.63−325225.56  &  TIC\,113636572  &$\times$  & 34.15  & 1(57.6\,\%) \\
    WD\,J214810.74−562613.14  &   TIC\,197765587 &$\times$  & 40.02  & 2(33.8\,\%) \\
    WD\,J220552.11−665934.73  &  TIC\,327712864  &$\times$  & 31.43  & 2(6.3\,\%) \\
    WD\,J223607.66−014059.65  &  TIC\,125250375  &$\times$  & 39.01  & 1(0.0\,\%) \\
    WD\,J235419.41−814104.96  &  TIC\,410255721  &$\times$  & 26.95  & 1(22.9\,\%) \\
    *WD\,J090212.89−394553.32  &  TIC\,191532802  & $\times$  & 36.41  & 1(36.8\,\%) \\
    *WD\,J200707.98−673442.18  & TIC 374346574 & $\times$  & 38.46  &1(20.9\,\%) \\

     \hline
    \end{tabular}
\end{table*}

\subsection{An eclipsing binary contaminating the {\it TESS} light curve}

\begin{figure}
\includegraphics[width=0.9\columnwidth]{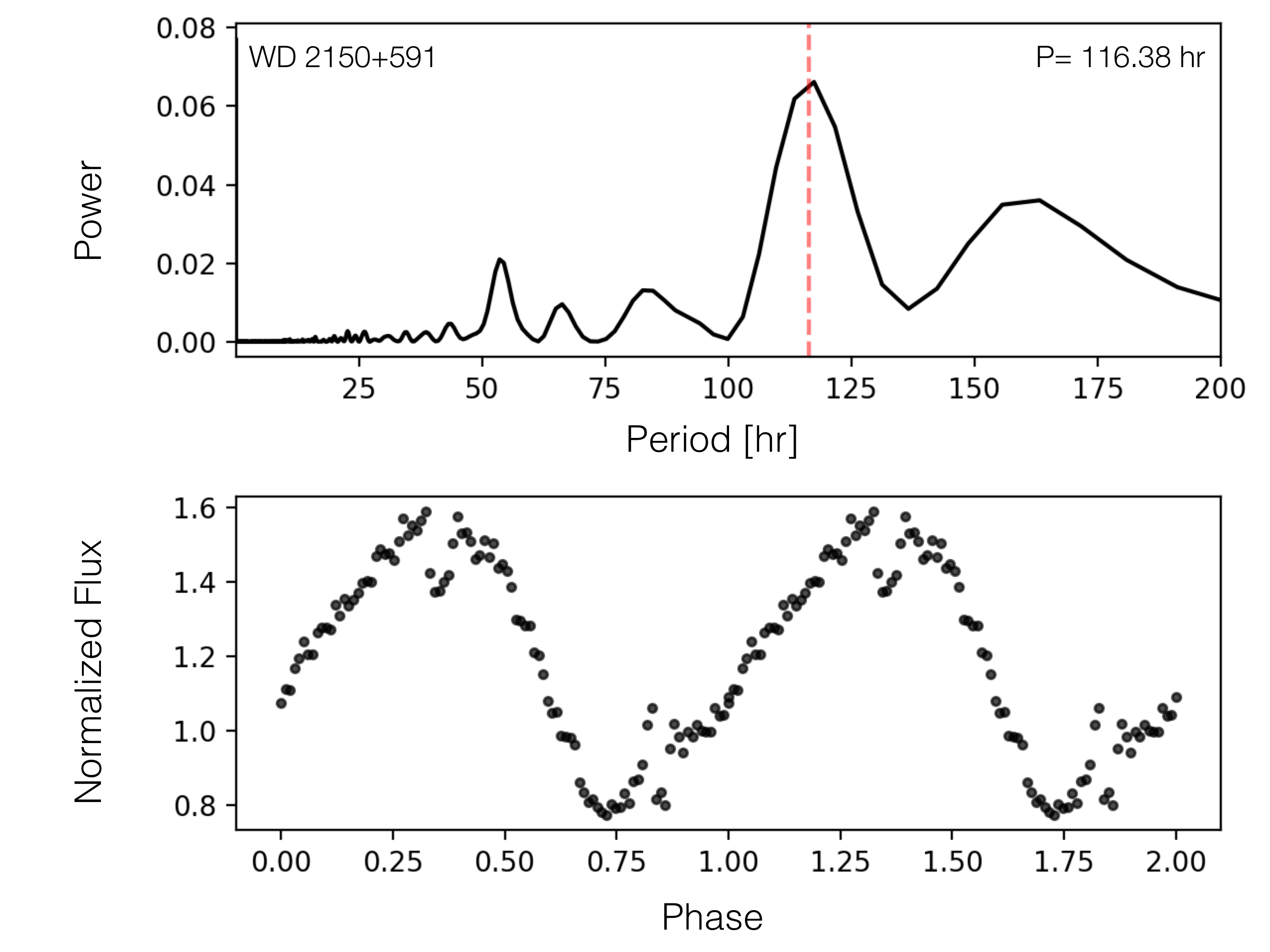}

    \caption{ Periodogram and phase-folded  {\it TESS} light curve of WD\,2150+591, the closest white dwarf in our sample (8.5\,pc). 
     We binned the phase-folded light curve (using 100 bins) to reduce the noise.
     This white dwarf was observed in five {\it TESS} sectors, by comparing the G-magnitude of each sector with the magnitude reported in the literature for this white dwarf, we found that two sectors are dominated by the light curve of a nearby eclipsing binary. Here we show the light curve corresponding to sector 56 with a 20 seconds cadence and a contamination level of 10.5\,\%. }
    \label{fig:false}
\end{figure}

During the {\it TESS} light curves analysis of WD\,2150+591, we noticed that the light curves of two sectors showed features resembling that of an eclipsing binary system. 
We modified the {\sc flux contamination tool} recently published by \citet{Schonhut23} with the aim to obtain the contamination level and G-magnitude of individual sectors in order to analyze this particular case. We compared the resulting G-magnitudes (by sector) with the white dwarf G-magnitude reported in literature. We found that the sectors showing the variations reminiscent of an eclipsing binary (Figure\,\ref{fig:false}) indeed were dominated by a nearby eclipsing binary star as the average G-magnitude was significantly smaller than that of the white dwarf. Therefore the variations one can find in the {\it TESS} database for WD\,2150+591 are clearly not produced by the white dwarf.

\subsection{Folding {\it TESS} data over the real rotation period}
\label{sec:second_peak}

\begin{figure*}
    \centering
    \includegraphics[width=1.9\columnwidth]{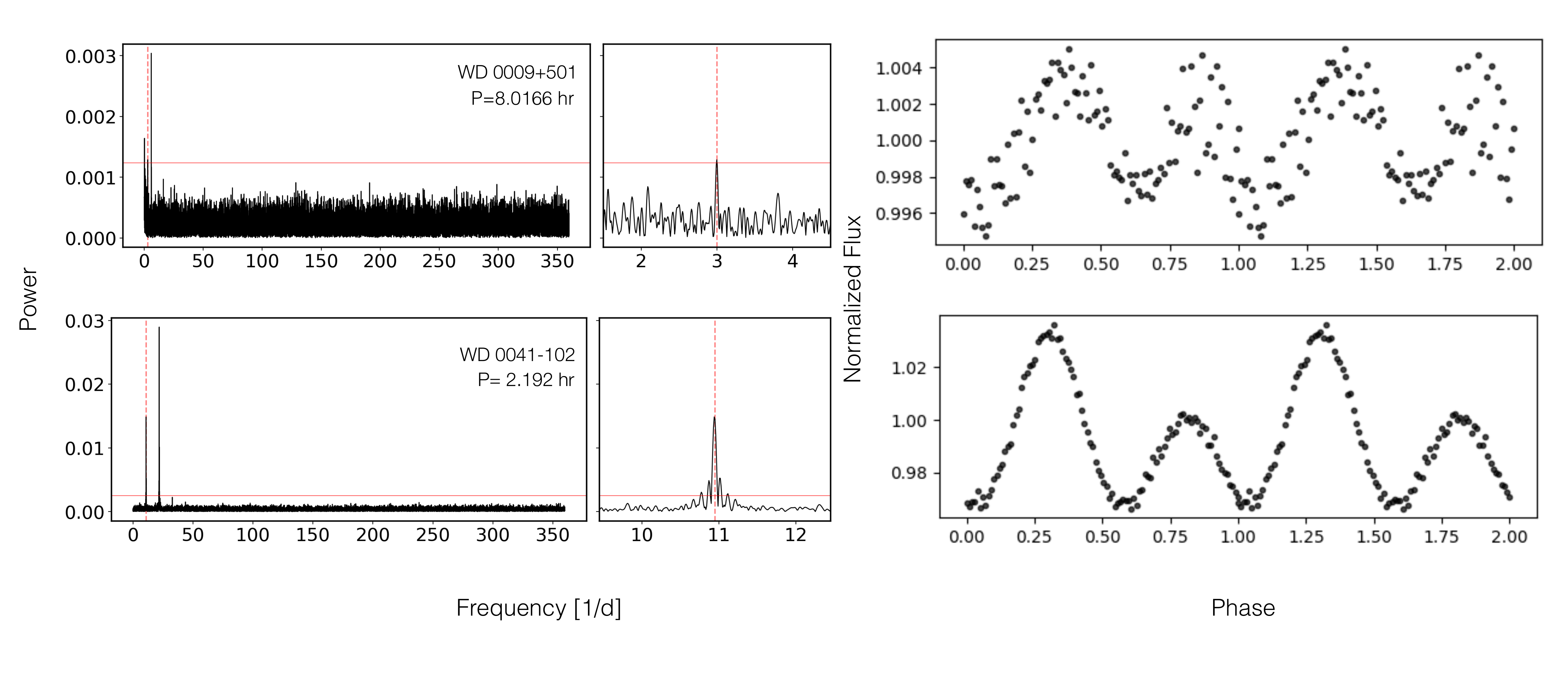}
    \caption{  Periodograms  (left: up to the Nyquist frequency; middle: zoomed in to the second highest peak) and phase-folded {\it TESS} light curves (right) of WD\,0009+501 and WD\,0041-102.
    The horizontal red line illustrates the false alarm probability level (FAP level) for a probability of $10^{-3}$. 
    For both objects the true rotation period is close to the second highest peak in the periodogram. The data of WD\,0041-102 was folded over the period indicated by the second highest peak in the periodogram ($2.192$\,hr) and we recover the double humped light curve previously reported by \citet{achilleosetal92-1}. 
    In the case of WD\,0009+501, the light curve is folded over a period of $8.016\pm0.083$\,hr, obtained by re-analyzing the full spectropolarimetry data set from \citet{valyavinetal05-1} supplemented with one all-night spectropolarimetry data set in the yet unpublished work of Bagnulo, Landstreet $\&$ Valyavin. This magnetic period is almost identical to twice the period of the highest peak in the periodogram ($8.017$\,hr) and close to the period indicated by the second highest peak in the periodogram ($8.007$\,hr). }
    
   \label{fig:TESS_LC_r}
\end{figure*}

For two white dwarfs in our sample, the strongest signal in the periodogram corresponds to half the orbital period. 
The independently measured period appears as significant second highest peak in the periodograms. 
In Fig.\ref{fig:TESS_LC_r} we show the {\it TESS} light curves folded over the true rotation period. In the case of WD\,0041-102, this period is identical to the period indicated by the second highest peak in the periodogram. 
For WD\,0009+501, however, the period corresponding to the second highest peak ($8.007$\,hr) provides a poor fit to the light curve and we instead used the slightly  different period ($8.016\pm0.083$\,hr) obtained by re-analyzing the full spectropolarimetry data set from \citet{valyavinetal05-1} supplemented with one all-night spectropolarimetry data set in the yet unpublished work of Bagnulo, Landstreet $\&$ Valyavin. The latter magnetic period provides a decent phase-folded light curve and is very close to the period corresponding to twice the period of the highest peak in the periodogram ($8.017$\,hr). 


\section{SPECULOOS observing log}

\begin{table}
    \centering
    \caption{ Observational specifications for targets observed with SPECULOOS. }
    \resizebox{8cm}{!}{\begin{tabular}{lllll}
    \hline 
        Name & Date & Filter & Exp. time & Telescope\\
          & [dd/mm/yyyy]  & & [s] &\\
    \hline
    WD\,2138-332 & 10-09-2021 & $g'$ & 60  & Callisto\\
     & 29-07-2022 & $g'$ & 60 & Europa\\
     & 30-07-2022 & $g'$ & 60 & Europa\\
     & 06-08-2022 & $g'$ & 60 & Ganymede\\
    LSPM\,J0107+2650 & 16-10-2021 & $g'$ & 1000& Europa\\
     & 17-10-2021 & $g'$ & 1000 & Europa\\
     & 29-07-2022 & $g'$ & 1000& Europa\\
     & 30-07-2022 & $g'$ & 1000& Europa\\
     & 06-08-2022 & $g'$ & 1000 & Ganymede\\
    \hline
    \end{tabular}}
    \label{tab:SPECULOOS}
\end{table}

\section{Rotation periods of magnetic and non-magnetic white dwarfs from the literature}

We carefully searched the literature to compile a list of previously measured rotation periods of magnetic (Table\,\ref{tab:brink}) and non-magnetic (Table\,\ref{tab:nomag}) white dwarfs. 

\begin{table*}
    \setlength{\tabcolsep}{5pt}
	\centering
	\caption{Rotation periods taken from the literature. We ignored white dwarfs for which only an estimate or a range of possible periods was given. The listed distances were obtained from the {\it Gaia} catalogue \citep{gaiaDistance21}, and some temperatures and masses were taken from \citet{Gentile19, Gentile21}. We assigned different abbreviations to identify the method used to measure the rotational period of a given white dwarf: Zeeman effect (Z), Polarimetry (P), Ground-based photometry (Ph) and Space-missions photometry (S).  White dwarf with spin periods measured with {\it TESS} are marked with and asterisk. }
	\label{tab:brink}
	\begin{tabular}{lllllllll} 
    \hline 
        Name & Period & Spec. & Temperature & Mass & Magnetic & Distance & Detection& Ref\\
         & [hr] & type& [K] & [M$_{\odot}]$& [MG] & [pc] &  &\\
    \hline
    WD\,0003-103 & $50.64 \pm 1.08$ & DQ & 19\,420 $\pm$ 920 & 1.14 & 1.47 & $150.69 \pm 3.22$ & Ph &12, 15\\
     *WD\,0009+501 & 8.0 & DAH & 6\,480 & 0.74 & 0.15-0.3 & 10.87 $\pm $ 0.01 & P & 38, 39\\ 
    *WD\,0011-134 & 0.74 & DAH & 5\,855 & 0.72 & 10 &  18.55 $\pm$ 0.01  & P &  31, 34, 49 \\ 
    *WD\,0041-102 & 2.18 & DBA & 20\,000 & 1.1 & 35 & $31.13 \pm 0.04$ & Z & 2, 3, 24\\
    WD\,0051+117 & 112.87 & DAH &20\,790 &  0.6& 0.25-0.37 & 115.67 $\pm$ 0.73  &  P   &  44\\
    WD\,0253+508 & $3.79 \pm 0.05$ & DAH & 15\,000 & 0.66 & 17 & $68.71 \pm 0.18$ & P &  1, 16, 36\\
    WD\,0316-849 & 0.0084 & DAH & 25\,970 & 1.26 & 185--425 & 29.38 $\pm$ 0.02 & Z & 4, 10 \\
    WD\,0553+053 & 0.45 & DAH & 5\,790 & 0.71 & 20 & $8.12 \pm 0.01$ &  Z & 3, 5, 9, 30, 41\\
    WD\,0637+477 & 39.16 & DAH & 13\,980 & 0.75 & 810-1070 & 40.20 $\pm$ 0.04 &  P  & 44 \\
    WD\,0756+437 & 6.68 & DC & 8\,500 & 1.07 & 300-337 & $21.95 \pm 0.02$ &  Ph   & 8, 30, 31\\
    *WD\,0912+536  &  31.93 & DC & 7\,170 & 0.74 & 100 & 10.27 $\pm$ 0.01& P   &  5, 15, 23, 49 \\
    WD\,1015+014 & 1.75 & DAH & 10\,000 & 0.91 & 50-90 & $49.39 \pm 0.16$ &  Ph  &8, 13, 36, 42\\
    WD\,1031+234 & 3.53 & DAH & 15\,000 & 0.93 & 500-1\,000 & $64.36 \pm 0.18$ &  Ph & 8, 35\\
    WD\,1045-091 & 2.75 & DAH & 9\,819 & 0.81 & 10-70 & 54.12 $\pm$ 0.21 & P  & 13\\
    WD\,1211-171 & 1.79 & DB & 12-23\,000 & 1.15 & 50 & $90.74 \pm 0.70$ &  Ph & 8, 33, 37\\
    WD\,1217+475  & 15.26 &  DAH   & 7\,500 $\pm$ 148 & 0.65 $\pm$ 0.02 & 18.5 $\pm$ 1.0 & 69.54 $\pm$ 0.38 & Z & 45 \\
    WD\,1312+098 & 5.42 & DAH & 20\,000 & 0.86 & 10 & $101.52 \pm 0.77$ &  P  & 36\\
    WD\,1346+384 & 0.67 & DAH & 30\,546 & 1.28 & 10 & 139.23$\pm$ 1.92 &  Ph  &20, 49 \\ 
    WD\,1533-057 & 1.89 & DAH & 20\,000 & 0.94 & 31 & $68.87 \pm 0.20$ &  P  & 1, 8, 25\\
    WD\,1639+537 & 1.93 & DAH & $7\,510 \pm 210$ & $0.67 \pm 0.07$ & 13 & $20.14 \pm 0.01$ &  Ph  &6, 14, 17, 40\\
    WD\,1743-520 & 68.16 & DAH & 20\,000 & 1.11 & 36 & $38.92 \pm 0.07$ &  P   &27, 40, 41\\
    WD\,1859+148 & 0.11 & DAH & 46\,000 & 1.34 & 800 & 41.40$\pm$ 0.08 &  Ph  & 11 \\
    WD\,1953-011 & 34.6 & DAH & $7\,920 \pm 200$ & $0.74 \pm 0.03$ & 0.1-0.5 & $11.57 \pm 0.01$ &  Ph  &5, 7, 28\\
    WD\,2047+372 & 5.83 & DAH& 14\,600 & 0.82 & 0.6 & 17.59$\pm$ 0.01 &   P  &21\\
    WD\,2051-208 & 1.42 & DAH & 21\,460& 1.24  & 0.3 & 31.27$\pm$ 0.03 &  P  & 44\\
                        WD\,2316+123 & 428.54 & DAH & $11\,000 \pm 1\,000$ & 0.86 & $45 \pm 5$ & $40.32 \pm 0.07$ &  Z  &26, 30, 36\\
    WD\,2329+267 & 2.76 & DAH & 11\,730 & 1.18 & 2.3 & $23.11 \pm 0.02$ &  P  &15, 29\\
    WD\,2254+076 & 0.37 & DAH & $13\,410 \pm 130$ & $1.30 \pm 0.01$ & 16.1 & 45.51 $\pm$ 0.21 &   Ph  &43\\
    WD\,2209+113 & 0.02 & DAH & $9\,021 \pm 160$ & 1.27 & 15 & 68.86 $\pm $ 1.54 &  Ph  &19\\
    *WD\,2359-434 & 2.69& DA & 8\,390& 0.83 &  0.1 & 8.33$\pm$ 0.01 & P &21\\
    Gaia\,DR3\,4479342339285057408 &  0.007 & DBA & 31\,200 & 1.33 & <1 & 75.55 $\pm$ 0.55 &   Ph  &32 \\
    SDSS\,J041926.91-011333.4 & 1.65 & DAHe & 7\,281.22 & 0.78 & 34.0 & 64.93 $\pm$ 0.45 & Z & 48\\
     SDSS\,J073227.97+662309.9 & 34.3 & DAHe & 8\,509.68 & 0.82 & 99.1 & 72.18 $\pm$ 0.44 & Z & 48\\
    SDSS\,J075224.18+472422.5 & 1.21 & DAHe & 7\,248.96 & 0.79 & 21.0 & 89.35 $\pm$ 1.24 & Z& 48 \\
    SDSS J075429.33+661105.7 & 1.37 & DAHe & 7\,274.33	 & 0.74 & 56.1 & 41.21 $\pm$ 0.09 & Z & 48\\
    SDSS\,J125230.93-023417.72 & 0.0881 &  DAEH  &  7\,856 $\pm$ 101 & 0.58 $\pm$ 0.03 & 5 $\pm$ 0.1 & 77.1 $\pm$ 0.7 & Z & 46\\
    WD\,J143019.29−562358.33 & 1.439 & DAHe & 8\,500 $\pm$ 170 & 0.83$\pm$ 0.03 &  5.8-8.9 & 66.33 $\pm$ 0.42 & Z & 47\\
    SDSS\,J150057.85+484002.3 & 1.42 & DAHe & 8\,225.29 & 0.79 & 19.0 & 115.21 $\pm$ 1.77 & Z& 48 \\
    SDSS\,J152934.98+292801.9 & 38.15 & DAH & $11\,920 \pm 190$ & 0.97 & 0.07 & 87.18 $\pm$ 0.60 &   Ph  & 18\\
    SDSS\,J161634.37+541011.4 & 1.59 & DAHe & 7\,937.51	 & 0.75 & 6.5 & 93.69 $\pm$ 0.86 & Z & 49\\
    LP\,705−64 & 1.21 & DAHe & 8\,440 $\pm $ 200 &  0.81 $\pm$ 0.04 & 9.4-22.2 &  52.66 $\pm$ 0.24 & Z & 47 \\
    ZTF\,J1901+1458 & 0.12 & DC2 & 46\,000 & 1.3 & 600-900 & 41.40 $\pm$ 0.08&   Ph   & 11\\
    \hline
	\end{tabular}
	\\
	1-\cite{ahilleos+wickramasinghe89-1};
2-\cite{achilleosetal92-1};
3-\cite{angel77-1};  
4-\cite{barstowetal95-1}; 
5-\cite{bergeronetal01-1}; 
6-\cite{brinkworthetal04-1}; 
7-\cite{brinkworthetal05-1}; 
8-\cite{brinkworthetal13-1}; 
9-\cite{bues+pragal89-1};  
10-\cite{burleighetal99-1}; 
11-\cite{caiazzoetal21-1}; 
12-\cite{dufouretal08-1}; 
13-\cite{euchneretal05-1}; 
14-\cite{ferrarioetal97-2}; 
15-\cite{ferrarioetal15-1}; 
16-\cite{friedrichetal97-1}; 
17-\cite{greensteinetal85-1}; 
18-\cite{kilicetal15-1};  
19-\cite{kilicetal21-1};  
20-\cite{Kuelebi09};   
21-\cite{Landstreet17};  
22-\cite{lawrieetal13-1};  
23-\cite{liebert76-1};   
24-\cite{liebertetal77-1};  
25-\cite{liebertetal85-1};   
26-\citet{Liebert85};   
27-\cite{martin+wickramasinghe78-1};  
28-\cite{maxtedetal00-1};   
29-\cite{moranetal98-1};  
30-\cite{putney+jordan95-1};  
31-\cite{putney97-1};   
32-\cite{Pshirkov20};   
33-\cite{reimersetal96-1};   
34-\cite{Bergeron92}; 
35-\cite{schmidtetal86-1};   
36-\cite{schmidt+norsworthy91-1};  
37-\cite{schmidtetal01-1};  
38-\cite{Valyavin05};   
39-\cite{Valyavin08};   
40-\cite{wegner77-1};   
41-\cite{wickramasinghe+martin79-1};   
42-\cite{wickramasinghe+cropper88-1};   
43-\cite{williamsetal22-1};   
44-Landstreet $\&$ Bagnulo (private communication);  
45-\cite{gaensicke20};  
46-\cite{Reding20};  
47-\cite{Reding23};  
48-\cite{Manser23};  
49-\cite{Lawrie13-2};  
50.-\cite{Angel71}.  

\end{table*}

\begin{table*}
    \setlength{\tabcolsep}{6pt}
	\centering
	\caption{Spin periods of non-magnetic white dwarfs measured using Kepler light curves of pulsating DA white dwarfs \citep{hermesetal17-1}. The distances were obtained from the {\it Gaia} catalogue \citep{gaiaDistance21}.}
	\label{tab:nomag}
	\begin{tabular}{lllllllll}
		\hline
        Name & Period & Spec. & Temperature & Mass & Distance & Ref\\
          & [hr] & type & [K] & [M$_{\odot}]$ & [pc]& \\
    \hline
    WD 0133-116 & 37.8 & DAV & 12\,300 & 0.62 & $32.74 \pm 0.26$ & 1\\
    WD 0415+271 & 52.8 & DAV & 11\,470 & 0.56 & $48.25 \pm 0.07$ & 2\\
    WD 0507+045.2 & 40.9 & DAV & 12\,010 & 0.72 & $49.46 \pm 0.08$ & 3\\
    WD 1137+423 & 5.9 & DAV & 11\,940 & 0.71 & $90.47 \pm 0.75$ & 4\\
    WD 1307+354 & 53.3 & DAV & 11\,120 & 0.65 & $48.01 \pm 0.07$ & 5\\
    WD 1349+552 & 41.8 & DAV & 12\,150 & 0.59 & $69.43 \pm 0.15$ & 6\\
    WD 1422+095 & 57.3 & DAV & 12\,220 & 0.67 & $33.34 \pm 0.02$ & 1\\
    WD 1425-811 & 13.0 & DAV & 12\,070 & 0.69 & $20.89 \pm 0.01$ & 7\\
    SDSS J1612+0830 & 1.9 & DAV & 11\,810 & 0.78 & $128.63 \pm 2.18$ & 8\\
    WD 1647+591 & 8.9 & DAV & 12\,510 & 0.83 & $10.94 \pm 0.02$ & 9\\
    WD 1935+276 & 14.5 & DAV & 12\,470 & 0.67 & $18.26 \pm 0.01$ & 10\\
    WD 0122+200 & 37.2 & DOV & 80\,000 & 0.53 & $609.38 \pm 29.7$ & 11\\
    WD 1159-034 & 33.6 & DOV & 140\,000 & 0.54 & $591.36 \pm 22.38$ & 12\\
    WD 2131+066 & 5.1 & DOV & 95\,000 & 0.55 & 107.25 $\pm$ 5.23 & 13\\
    KIC 8626021 & 44.6 & DBV & 30\,000 & 0.59 & $377.92 \pm 16.0$ & 14\\
    WD 0112+104 & 10.2 & DBV & 31\,000 & 0.52 & $110.92 \pm 0.56$ & 15\\
    KIC 4357037 & 22.0 & DAV & 12\,650 & 0.62 & $206.31 \pm 4.93$ & 16\\
    KIC 4552982 & 18.4 & DAV & 10\,950 & 0.67 & $132.15 \pm 1.36$ & 17\\
    KIC 7594781 & 26.8 & DAV & 11\,730 & 0.67 & $172.44 \pm 2.73$ & 16\\
    KIC 10132702 & 11.2 & DAV & 11\,940 & 0.68 & $254.77 \pm 10.71$ & 16\\
    KIC 11911480 & 74.7 & DAV & 11\,580 & 0.58 & $181.88 \pm 2.87$ & 18\\
    KIC 60017836 & 6.9 & DAV & 10\,980 & 0.57 & $18.66 \pm 0.01$ & 16\\
    EPIC 201719578 & 26.8 & DAV & 11\,070 & 0.57 & $169.80 \pm 4.96$ & 16\\
    EPIC 201730811 & 2.6 & DAV & 12\,480 & 0.58 & $130.99 \pm 1.32$ & 19\\
    EPIC 201802933 & 31.3 & DAV & 12\,330 & 0.68 & $144.63 \pm 2.71$ & 16\\
    EPIC 201806008 & 31.3 & DAv & 10\,910 & 0.61 & $40.31 \pm 0.06$ & 16\\
    EPIC 210397465 & 49.1 & DAV & 11\,200 & 0.45 & $168.26 \pm 3.14$ & 16\\
    EPIC 211596649 & 81.8 & DAV & 11\,600 & 0.56 & $258.13 \pm 18.32$ & 16\\
    EPIC 211629697 & 64.0 & DAV & 10\,600 & 0.48 & $199.24 \pm 7.26$ & 16\\
    EPIC 211914185 & 1.1 & DAV & 13\,590 & 0.88 & $207.46 \pm 12.82$ & 20\\
    EPIC 211926430 & 25.4 & DAV & 11\,420 & 0.59 & $154.91 \pm 2.9$ & 16\\
    EPIC 228682478 & 109.1 & DAV & 12\,070 & 0.72 & $174.00 \pm 4.66$ & 16\\
    EPIC 229227292 & 29.4 & DAV & 11\,210 & 0.62 & $88.66 \pm 0.67$ & 16\\
    EPIC 220204626 & 24.3 & DAV & 11\,620 & 0.71 & $88.66 \pm 0.67$ & 16\\
    EPIC 220258806 & 30.0 & DAV & 12\,800 & 0.66 & $81.57 \pm 0.37$ & 16\\
    EPIC 220347759 & 31.7 & DAV & 12\,770 & 0.66 & $165.48 \pm 2.91$ & 16\\
	\hline
	\end{tabular}
    \\
    1-\cite{giammicheleetal16-1}; 2-\cite{dolezetal16-1}; 3-\cite{fuetal13-1}; 4-\cite{suetal14-1}; 5-\cite{pfeifferetal96-1};\\ 6-\cite{bognaretal16-1}; 7-\cite{bradley01-1}; 8-\cite{castanheiraetal13-1}; 9-\cite{kepleretal95-1}; 10-\cite{pech-vauclair06-1};\\ 11-\cite{fuetal07-1}; 12-\cite{charpinetetal09-1}; 13-\cite{kawaleretal95-1}; 14-\cite{ostensenetal11-1}; 15-\cite{hermesetal17-2};\\ 16-\cite{hermesetal17-1}; 17-\cite{belletal15-1}; 18-\cite{greissetal14-1}; 19-\cite{hermesetal15-1}; 20-\cite{hermesetal17-3}
\end{table*}

\bsp	
\label{lastpage}
\end{document}